\documentclass[aps, prd, floatfix, nofootinbib, showpacs, twocolumn, superscriptaddress]{revtex4-1}

\usepackage{amsmath}
\usepackage{amssymb}
\usepackage{graphicx}
\usepackage{color}
\usepackage{xspace}

\usepackage[mathscr,scaled=1.15]{urwchancal}
\DeclareFontFamily{OT1}{pzc}{}
\DeclareFontShape{OT1}{pzc}{m}{it}%
{<-> s * [1.15] pzcmi7t}{}
\DeclareMathAlphabet{\mathpzc}{OT1}{pzc}{m}{it}

\definecolor{purple}{rgb}{0.5,0,0.5}
\definecolor{blue}{rgb}{0.0,0,0.9}
\definecolor{prdblue}{rgb}{0.133,0.118,0.498}
\usepackage[colorlinks=true, pdfstartview=FitV, linkcolor=prdblue, citecolor= prdblue, urlcolor=prdblue]{hyperref}

\begin{document}

\title{Poincar\'e-covariant analysis of heavy-quark baryons}

\author{Si-Xue~Qin}
\email[]{sqin@cqu.edu.cn}
\affiliation{Department of Physics, Chongqing University, Chongqing 401331, P.R. China}

\author{Craig D.~Roberts}
\email[]{cdroberts@anl.gov}
\affiliation{Physics Division, Argonne National Laboratory, Argonne,
Illinois 60439, USA}

\author{Sebastian M. Schmidt}
\email[]{s.schmidt@fz-juelich.de}
\affiliation{
Institute for Advanced Simulation, Forschungszentrum J\"ulich and JARA, D-52425 J\"ulich, Germany}

\date{26 January 2018}

\begin{abstract}
We use a symmetry-preserving truncation of meson and baryon bound-state equations in quantum field theory in order to develop a unified description of systems constituted from light- and heavy-quarks.  In particular, we compute the spectrum and leptonic decay constants of ground-state pseudoscalar- and vector-mesons: $q^\prime \bar q$, $Q^\prime \bar Q$, with $q^\prime,q=u,d,s$ and $Q^\prime,Q = c,b$; and the masses of $J^P=3/2^+$ baryons and their first positive-parity excitations, including those containing one or more heavy quarks.
This Poincar\'e-covariant analysis predicts that such baryons have a complicated angular momentum structure.  For instance, the ground states are all primarily $S$-wave in character, but each possesses $P$-, $D$- and $F$-wave components, with the $P$-wave fraction being large in the $qqq$ states; and the first positive-parity excitation in each channel has a large $D$-wave component, which grows with increasing current-quark mass, but also exhibits features consistent with a radial excitation.
The configuration space extent of all such baryons decreases as the mass of the valence-quark constituents increases.
\end{abstract}

\maketitle


\section{Introduction}
Constituent-quark models have long predicted the existence of doubly- and triply-heavy baryons \cite{Roberts:2007ni, Crede:2013sze, Chen:2016spr}.  Some of those predictions have recently been qualitatively confirmed via numerical simulations of lattice-regularised QCD \cite{Brown:2014ena, Padmanath:2015jea, Bali:2015lka}.  Within the models the level ordering and splittings are sensitive to the shape assumed for the potential between the constituent quarks, so there are some quantitative disagreements.  Such problems might be ameliorated though the use of a potential constructed systematically using an effective field theory framework \cite{LlanesEstrada:2011kc, Brambilla:2013vx}, but issues of convergence may obscure this path.

Notwithstanding such theoretical issues, the experimental search for these states is underway.  There is empirical evidence for the existence of baryons containing two heavy quarks \cite{Mattson:2002vu, Aaij:2015tga, Aaij:2017ueg}; and although triply-heavy baryons have not yet been seen, it has been argued that even states constituted from three valence $b$-quarks could be produced at the large hadron collider \cite{Wu:2012wj}.

The beauty of triply-heavy baryons is the analogies one might draw between these systems and heavy quarkonia states, about which much is known \cite{Brambilla:2010cs}; the curse is that any continuum bound-state treatment must involve a strategy for producing an accurate solution to the three-body problem.  The three-body challenge has long been faced in connection with quantum-mechanical potential-models and strategies exist \cite{Capstick:2000qj, Giannini:2015zia}.   In relativistic quantum field theory, on the other hand, the three-valence-body problem is a greater challenge.   Tackling it is made worthwhile because one can formulate the problem in such a way as to maintain a traceable connection with QCD, making predictions that can systematically be improved; and also unify the treatment of mesons and baryons, and light- and heavy-quark composites within a single framework.

A contemporary perspective on the continuum bound-state problem in QCD is provided, \emph{e.g}.\ in Refs.\,\cite{Chang:2011vu, Bashir:2012fs, Roberts:2015lja, Horn:2016rip, Eichmann:2016yit}; and although a treatment of (heavy-heavy) mesons will be an important part of the analysis herein, the essentially new element is solution of a Faddeev equation for triply-heavy baryons.  This approach to baryons was introduced in Refs.\,\cite{Cahill:1988dx, Burden:1988dt, Reinhardt:1989rw, Efimov:1990uz}, which capitalised on the role of diquark correlations in order to simplify the problem \cite{Cahill:1987qr, Maris:2002yu, Maris:2004bp, Segovia:2015ufa}; but we will adapt the formulation in Ref.\,\cite{Eichmann:2009qa} and solve the three-valence-body problem directly, under the assumption that two-body interactions dominate in forming a baryon bound-state.

We describe the Faddeev equation in Sec.\,\ref{SecFaddeevE}, focusing on $J^P=3/2^+$ states in this first Poincar\'e-covariant study of triply-heavy baryons.  Sec.\,\ref{SeccalG} details the quark-quark interaction that we use in solving for all elements relevant to our two- and three-valence-body bound-state equations.  In Sec.\,\ref{SecMeson} we report a range of properties of light-quark mesons and baryons, which provide a benchmark for our results in the heavy-quark sector -- mesons, Sec.\,\ref{SecMesonHeavy}; and baryons, Sec.\,\ref{SecQQQ}.  Sec.\,\ref{SecEpilogue} provides a summary and perspective.

\section{Three-body Faddeev Equation}
\label{SecFaddeevE}
The Faddeev amplitude for a $J=3/2$ baryon can be written in terms of a Rarita-Schwinger spinor as follows:
\begin{align}
\nonumber
\mathbf{\Psi}^{\alpha_1\alpha_2\alpha_3,\delta}_{\mu; c_1 c_2 c_3}&(p_1,p_2,p_3;P)  \nonumber \\
& = \tfrac{1}{\surd 6} \varepsilon_{c_1 c_2 c_3} {\Psi}^{\alpha_1\alpha_2\alpha_3,\delta}_{\mu}(p_1,p_2,p_3;P)  \,,
\label{FaddeevAmp}
\end{align}
where $\mu$ is the spinor's Lorentz index;
$\alpha_{1,2,3}$, $\delta$ are spinor indices for the three valence quarks and baryon, respectively;
$c_{1,2,3}$ are colour indices;
and $P=p_1+p_2+p_3$, $P^2 = -M_{\rm baryon}^2$, where $p_{1,2,3}$ are the valence-quark momenta.  Herein we only solve directly for $QQQ$ baryons, so the flavour structure is trivial; and owing to explicit antisymmetry under the exchange of colour indices, $\Psi_\mu$ is completely symmetric under exchange of any other pair of valence-quark labels.

\begin{figure}[!t]
\begin{center}
\begin{tabular}{lcr}
\parbox[c]{0.33\linewidth}{\includegraphics[clip,width=\linewidth]{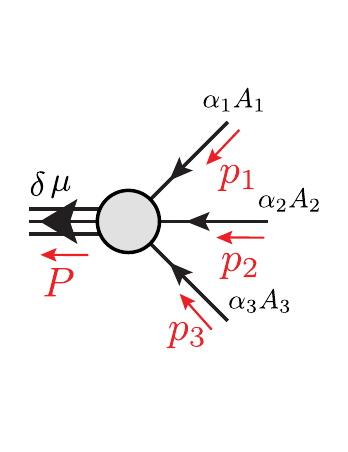}} &
 \hspace*{-1em} \mbox{\large $= \sum_{\{1,2,3\}}$} \hspace*{-0.5em} &
\parbox[c]{0.47\linewidth}{\includegraphics[clip,width=\linewidth]{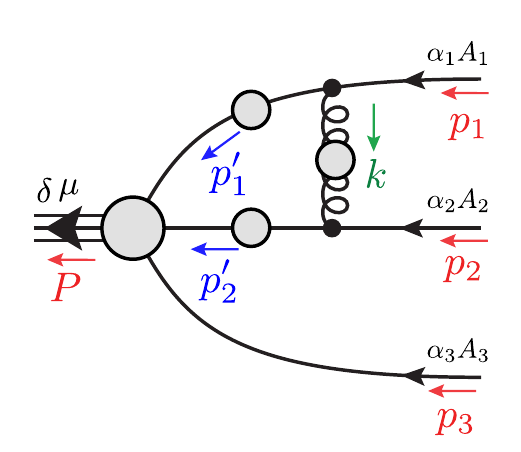}}\vspace*{-0ex}
\end{tabular}
\end{center}
\caption{\label{FEimage}
Three-body Faddeev equation in Eq.\,\eqref{eq:faddeev0}, solved herein for the mass and amplitude of $qqq$ and $QQQ$ baryons, $q=u,d,s$, $Q=c,b$.
Amplitude:  vertex on the lhs;
spring with shaded circle: quark-quark interaction kernel in Eq.\,\eqref{KDinteraction};
and solid line with shaded circle: dressed-propagators for scattering quarks, obtained by solving a gap equation with the same interaction.
}
\end{figure}

The continuum bound-state problem is defined by a collection of coupled integral equations.  A tractable system of equations is only obtained once a truncation scheme is specified; and a systematic, symmetry-preserving approach is described in Refs.\,\cite{Munczek:1994zz, Bender:1996bb, Binosi:2016rxzd}.  The leading-order term is the widely-used rainbow-ladder (RL) truncation, which is known to be accurate for ground-state light-quark vector- and isospin-nonzero-pseudoscalar-mesons, and related ground-state octet and decouplet baryons \cite{Chang:2011vu, Bashir:2012fs, Roberts:2015lja, Horn:2016rip, Eichmann:2016yit}; and, with judicious modification, heavy-heavy $S$-wave quarkonia \cite{Ding:2015rkn}.  RL truncation is accurate in these channels because corrections largely cancel owing to the preservation of relevant WGT identities ensured by the scheme \cite{Munczek:1994zz, Bender:1996bb, Binosi:2016rxzd}.  Consequently, it should be equally reliable for ground-state $QQQ$ baryons; and, hence, we consider the following Faddeev equation,  depicted in Fig.\,\ref{FEimage}:
\begin{subequations}
 \label{eq:faddeev0}
\begin{align}
{\Psi}^{\alpha_1\alpha_2\alpha_3,\delta}_{\mu}&(p_1,p_2,p_3)
= \sum_{j=1,2,3} \big[ {\mathscr K} S S \Psi_\mu \big]_j\,, \\
\big[ {\mathscr K} S S \Psi_\mu \big]_3 & =
\int_{dk} \mathscr{K}_{\alpha_1\alpha_1',\alpha_2\alpha_2'}(p_1,p_2;p_1',p_2') \nonumber\\
& \rule{-4em}{0ex} \times S_{\alpha_1'\alpha_2''}(p_1') S_{\alpha_2'\alpha_2''}(p_2') {\Psi}^{\alpha_1''\alpha_2''\alpha_3;\delta}_{\mu}(p_1',p_2',p_3) \,,
\end{align}
\end{subequations}
where $\int_{dk}$ represents a translationally-invariant definition of the four-dimensional integral, $\big[ {\mathscr K} S S \Psi_\mu \big]_{1,2}$ are obtained from $\big[ {\mathscr K} S S \Psi_\mu \big]_3$ by cyclic permutation of indices, and additional details concerning the structure of the Poincar\'e-covariant Faddeev amplitude are supplied in Appendix\,\ref{FAmplitude}.

\section{Quark-quark interaction}
\label{SeccalG}
The key element in all analyses of the continuum bound-state problem for hadrons is the quark-quark scattering kernel.  In RL truncation that can be written ($k = p_1-p_1^\prime = p_2^\prime -p_2$):
\begin{subequations}
\label{KDinteraction}
\begin{align}
\mathscr{K}_{\alpha_1\alpha_1',\alpha_2\alpha_2'} & = {\mathpzc G}_{\mu\nu}(k) [i\gamma_\mu]_{\alpha_1\alpha_1'} [i\gamma_\nu]_{\alpha_2\alpha_2'}\,,\\
 {\mathpzc G}_{\mu\nu}(k) & = \tilde{\mathpzc G}(k^2) T_{\mu\nu}(k)\,,
\end{align}
\end{subequations}
where $k^2 T_{\mu\nu}(k) = k^2 \delta_{\mu\nu} - k_\mu k_\nu$.
Thus, in order to define all elements in Eq.\,\eqref{eq:faddeev0} and hence the bound-state problem, it remains only to specify $\tilde{\mathpzc G}$; and following two decades of study, the following form has been found appropriate for RL truncation \cite{Qin:2011dd, Qin:2011xq} ($s=k^2$):
\begin{align}
\label{defcalG}
 \tfrac{1}{Z_2^2}\tilde{\mathpzc G}(s) & =
 \frac{8\pi^2}{\omega^4} D e^{-s/\omega^2} + \frac{8\pi^2 \gamma_m \mathcal{F}(s)}{\ln\big[ \tau+(1+s/\Lambda_{\rm QCD}^2)^2 \big]}\,,
\end{align}
where $\gamma_m=12/(33-2N_f)$, with $N_f=5$ herein; $\tau=e^2-1$; $\mathcal{F}(s)=(1-e^{-s/[4m_t^2]})/s$, with $m_t=0.5\,$GeV; and $Z_2$ is the dressed-quark wave function renormalisation constant.  We employ a momentum-subtraction scheme in renormalising gap and inhomogeneous vertex equations.  Hence, for QCD with five active quark flavours, wherewith  \cite{Olive:2016xmw} $\Lambda^{\rm \overline{MS}}_{\rm QCD} = 0.21\,$GeV, in Eq\,\eqref{defcalG} we use \cite{Boucaud:2008gn}:
\begin{eqnarray}
\label{LambdaQCD}
	\Lambda^{\rm \overline{MOM}}_{\rm QCD} = \Lambda^{\rm \overline{MS}}_{\rm QCD} e^{\frac{507-40N_f}{792-48N_f}} = 0.36\,{\rm GeV.}
\end{eqnarray}

The development of Eqs.\,\eqref{KDinteraction}, \eqref{defcalG} is summarised in Ref.\,\cite{Qin:2011dd} and their connection with QCD is described in Ref.\,\cite{Binosi:2014aea}; but it is worth reiterating some of that material.

The interaction in Eqs.\,\eqref{KDinteraction}, \eqref{defcalG} is deliberately consistent with that determined in studies of QCD's gauge sector, which indicate that the gluon propagator is a bounded, regular function of spacelike momenta that achieves its maximum value on this domain at $k^2=0$ \cite{Bowman:2004jm, Boucaud:2011ug, Ayala:2012pb, Aguilar:2012rz, Binosi:2016xxu, Binosi:2016nme}, and the dressed-quark-gluon vertex does not possess any structure which can qualitatively alter these features \cite{Skullerud:2003qu, Bhagwat:2004kj}.  It also preserves the one-loop renormalisation group behaviour of QCD so that, \emph{e.g}.\ the quark mass-functions one obtains are independent of the renormalisation point.  On the other hand, in the infrared, \emph{i.e}.\ $k^2 \lesssim \Lambda_{\rm QCD}^2$, Eq.\,\eqref{defcalG} defines a two-parameter model, the details of which determine whether confinement and/or dynamical chiral symmetry breaking (DCSB) are realised in solutions of the dressed-quark gap equations.

Computations \cite{Qin:2011dd, Qin:2011xq} reveal that observable properties of light-quark ground-state vector- and isospin-nonzero pseudoscalar-mesons are practically insensitive to variations of $\omega \in [0.4,0.6]\,$GeV, so long as
\begin{equation}
 \varsigma^3 := D\omega = {\rm constant}.
\label{Dwconstant}
\end{equation}
This feature also extends to numerous properties of the nucleon and $\Delta$-baryon \cite{Eichmann:2008ef, Eichmann:2012zz}.  The value of $\varsigma$ is chosen so as to obtain the measured value of the pion's leptonic decay constant, $f_\pi$; and in RL truncation this requires ($q=u,d,s$)
\begin{equation}
\label{varsigmalight}
\varsigma_{q} =0.80\,{\rm GeV.}
\end{equation}

Another way of looking at Eq.\,\eqref{defcalG} is suggested by Refs.\,\cite{Binosi:2014aea, Binosi:2016nme}.  Namely, one can sketch a connection with QCD's renormalisation-group-invariant process-independent effective charge by writing
\begin{equation}
\label{alphaRL}
\tfrac{1}{4\pi}\tilde{\mathpzc G}(s) \approx \frac{\tilde\alpha_{\rm PI}(s)}{s + \tilde m_g^2(s)}\,,\;
 m_g^2(s) = \frac{\tilde m_0^4}{s + \tilde m_0^2}\,,
\end{equation}
and extract $\tilde\alpha_{\rm PI}(0)=:\tilde\alpha_0$, $\tilde m_0$ via a least-squares fit on an infrared domain: $s\lesssim (4\Lambda_{\rm QCD})^2$.  In this way, one obtains
\begin{equation}
\tfrac{1}{\pi}\tilde\alpha_0^{\rm RL} = 9.7\,,\; \tilde m_0^{\rm RL} = 0.54\,{\rm GeV}\,,\;
\end{equation}
$\alpha_0^{\rm RL}/\pi/[m_0^{\rm RL}]^2 \approx 33\,$GeV$^{-2}$.  Comparison of these values with those predicted via a combination of continuum and lattice analyses of QCD's gauge sector \cite{Binosi:2016nme}: $\alpha_0/\pi \approx 0.95$, $m_0 \approx 0.5\,$GeV, $\alpha_0/\pi/m_0^2 \approx 4.2\,$GeV$^{-2}$, confirms an earlier observation \cite{Binosi:2014aea} that the RL interaction defined by Eqs.\,\eqref{KDinteraction}, \eqref{defcalG} has roughly the correct shape, but is an order-of-magnitude too large in the infrared.  As explained elsewhere \cite{Chang:2009zb, Chang:2010hb, Chang:2011ei}, this is because Eq.\,\eqref{KDinteraction} suppresses all effects associated with DCSB in bound-state equations \emph{except} those expressed in $\tilde{\mathpzc G}(k^2)$, and therefore a description of hadronic phenomena can only be achieved by overmagnifying the gauge-sector interaction strength at infrared momenta.


Our primary foci herein are systems involving heavy-quarks, so it is pertinent to remark that RL truncation has also been explored in connection with heavy-light mesons and heavy-quarkonia \cite{Bhagwat:2004hn, Rojas:2014aka, Hilger:2014nma, Ding:2015rkn}.  Those studies reveal that improvements to RL are critical in heavy-light systems; and an interaction strength for the RL kernel fitted to pion properties alone is not optimal in the treatment of heavy quarkonia.  Both observations are readily understood, but we focus on the latter because it is most relevant to our study.

Recall, therefore, that for meson bound-states it is now possible \cite{Chang:2009zb, Chang:2010hb, Chang:2011ei} to employ sophisticated kernels which overcome many of the weaknesses of RL truncation.  The new technique is symmetry preserving and has an additional strength, \emph{i.e}.\ the capacity to express DCSB nonperturbatively in the integral equations connected with bound-states.  Owing to this feature, the scheme is described as the ``DCSB-improved'' or ``DB'' truncation.  In a realistic DB truncation, $\varsigma^{\rm DB} \approx 0.6\,$GeV; a value which coincides with that predicted by solutions of QCD's gauge-sector gap equations \cite{Binosi:2014aea, Binosi:2016wcx, Binosi:2016nme}.
Straightforward analysis shows that corrections to RL truncation largely vanish in the heavy+heavy-quark limit;
%
%
hence the aforementioned agreement entails that RL truncation should provide a reasonable approximation for systems involving only heavy-quarks so long as one employs $\varsigma^{\rm DB}$ as the infrared mass-scale.  In heavy-quark systems we therefore employ Eqs.\,\eqref{KDinteraction}, \eqref{defcalG} as obtained using
\begin{equation}
\label{varsigmaQ}
\varsigma_Q = 0.6\,{\rm GeV}\,.
\end{equation}

\section{Light Quarks}
\label{SecMeson}
%
%
In order to compute properties of systems constituted from lighter valence-quarks, $q=u,d,s$, all that remain to be specified are the Higgs-generated current-quark masses, $m_{q}$.  We work in the isospin symmetric limit: $m_u=m_d$, and employ a mass-independent momentum-subtraction renormalisation scheme at a far-ultraviolet scale $\zeta_{19} = 19\,$GeV, wherewith the choices
\begin{equation}
\label{currentuds}
m_{u,d}^{\zeta_{19}} = 3.3\,{\rm MeV}\,,\; m_s^{\zeta_{19}} = 74.6\,{\rm MeV}\,,
\end{equation}
when used to specify the gap equation solutions that feed into the Bethe-Salpeter and Faddeev equations, yield the results in Table\,\ref{obsuds}.\footnote{We reiterate here that the mass-scale in Eq.\,\eqref{varsigmalight} makes no allowance for the effect of corrections to RL truncation on light-hadron observables.  This issue is canvassed elsewhere \cite{Eichmann:2008ae}, with the conclusion that for systems in which orbital angular momentum does not play a big role, the impact of such corrections may largely be absorbed in a redefinition of this scale; something we discussed in connection with Eq.\,\eqref{varsigmaQ}.}
(Aspects of our approach to solving the Faddeev equation are detailed in Appendix~\ref{AppNumerical}.)
The values in Eq.\,\eqref{currentuds} correspond to renormalisation-group-invariant masses $\hat m_{u,d}=6.3\,$MeV, $\hat m_s=146\,$MeV; one-loop-evolved masses at 2\,GeV of
\begin{equation}
m_{u=d}^{2\,{\rm GeV}}=4.8\,{\rm MeV}\,, \; m_{s}^{2\,{\rm GeV}}=110\,{\rm MeV}\,;
\end{equation}
Euclidean constituent quark masses
\begin{equation}
M^E_{u,d}=0.41\,{\rm GeV},\quad M^E_s=0.57\,{\rm GeV},
\end{equation}
defined via $M_q^E = \{k|M_q(k)=k\}$, where $M_q(k)$ is the nonperturbative solution of the appropriate gap equation;
and give $\hat m_s/\hat m_{u=d}=23$.
They are consequently compatible with modern estimates by other means \cite{Olive:2016xmw}.

\begin{table}[t]
\caption{\label{obsuds}
Computed values for a range of light-quark-hadron properties, obtained using the quark-quark scattering kernel described in Sec.\,\ref{SeccalG} to specify the relevant gap-, Bethe-Salpeter- and Faddeev-equations: $\Delta^\prime$ and $\Omega^\prime$ denote the first positive-parity excitations in these channels.  The interaction scale is stated in Eq.\,\eqref{varsigmalight} and the current-quark masses in Eq.\,\eqref{currentuds} were chosen to reproduce the empirical values of $m_\pi=0.14\,$GeV, $m_K=0.50\,$GeV.  ($Z_2(\zeta_{19})\approx 1$.)  The rms relative-error per degree-of-freedom between calculation and experiment is 7\%.
\emph{N.B}.\ The results in the rightmost four columns of panel~B were obtained using the equal spacing rule described in connection with Eqs.\,\eqref{eqESR}.
(All quantities listed in GeV; and where known, experimental values drawn from Ref.\,\cite{Olive:2016xmw}.)}
\begin{center}
\begin{tabular*}
{\hsize}
{l@{\extracolsep{0ptplus1fil}}
|l@{\extracolsep{0ptplus1fil}}
l@{\extracolsep{0ptplus1fil}}
l@{\extracolsep{0ptplus1fil}}
l@{\extracolsep{0ptplus1fil}}
l@{\extracolsep{0ptplus1fil}}
l@{\extracolsep{0ptplus1fil}}
l@{\extracolsep{0ptplus1fil}}
l@{\extracolsep{0ptplus1fil}}}\hline
(A)\;       & $f_\pi$ & $f_K$ & $m_\rho$ & $f_\rho$ & $m_{K^\ast}$ & $f_{K^\ast}$ & $m_\phi$ & $f_\phi$   \\
herein\; & $0.094$ & $0.11$ & $0.75$ & $0.15$ & $0.95$ & $0.18$ & $1.09$ & $0.19$ \\
expt. & $0.092$ & $0.11$ & $0.78$ & $0.15$ & $0.89$  & $0.16$ & $1.02$ & $0.17$ \\\hline
\end{tabular*}
\smallskip

\begin{tabular*}
{\hsize}
{l@{\extracolsep{0ptplus1fil}}
|l@{\extracolsep{0ptplus1fil}}
l@{\extracolsep{0ptplus1fil}}
l@{\extracolsep{0ptplus1fil}}
l@{\extracolsep{0ptplus1fil}}
|l@{\extracolsep{0ptplus1fil}}
l@{\extracolsep{0ptplus1fil}}
l@{\extracolsep{0ptplus1fil}}
l@{\extracolsep{0ptplus1fil}}}\hline
(B)\;       & $m_{\Delta}$ & $m_{\Delta^\prime}$ & $m_{\Omega}$ & $m_{\Omega^\prime}$ \;
    & $m_{\Sigma^\ast}$ & $m_{\Xi^\ast}$ & $m_{\Sigma^{\ast\prime}}$ & $m_{\Xi^{\ast\prime}} $
    \\
herein\;  & $1.21$ & $1.46$ & $1.67$ & 1.96 & 1.36 & 1.52 & 1.63 & 1.79 \\
expt. & $1.21$ & $1.51$ & $1.67$ & $-$ & 1.38 & 1.53 & $-$ & $-$ \\\hline
\end{tabular*}
\end{center}
\end{table}

It is worth making a few remarks here.  First recall the equal spacing rule \cite{Okubo:1961jc, GellMann:1962xb}:
\begin{subequations}
\label{eqESR}
\begin{align}
1.38 = m_{\Sigma^\ast} & \approx m_{\Sigma^\ast}^{\rm interp} := \tfrac{2}{3} m_\Delta +  \tfrac{1}{3} m_{\Omega}= 1.36\,,\\
1.53 = m_{\Xi^\ast} & \approx m_{\Xi^\ast}^{\rm interp} := \tfrac{1}{3} m_\Delta +  \tfrac{2}{3} m_{\Omega} = 1.52\,,
\end{align}
\end{subequations}
where the listed values are empirical (GeV).  This approximate linear growth of mass with the infrared scale of the dressed-masses of a baryon's constituents is preserved in RL truncation treatments of the Faddeev equation \cite{Roberts:2011cf, Eichmann:2016hgl, Lu:2017cln}.  Empirically, similar correspondences are found in the light-quark sector for the masses and decay constants of vector mesons (GeV):
\begin{subequations}
\begin{align}
0.89 = m_{K^\ast} & \approx m_{K^\ast}^{\rm interp} := \tfrac{1}{2} m_\rho + \tfrac{1}{2} m_\phi = 0.90\,,\\
0.16 = f_{K^\ast} & \approx f_{K^\ast}^{\rm interp} := \tfrac{1}{2} f_\rho + \tfrac{1}{2} f_\phi = 0.16\,.
\end{align}
\end{subequations}
Again, such near-linear evolution is found for these systems in RL truncation \cite{Krassnigg:2004if}, which also predicts the same behaviour for the leptonic decay constants of light-quark isospin-nonzero pseudoscalar mesons \cite{Krassnigg:2004if} (GeV):
\begin{equation}
\label{eqfKESR}
0.11 = f_K \approx f_{K}^{\rm interp} := \tfrac{1}{2} f_\pi + \tfrac{1}{2} f_{s^\prime \bar s} = 0.11\,,
\end{equation}
where $f_{s^\prime \bar s}=0.13\,$GeV is a RL prediction for the leptonic decay constant of a fictitious ``heavy-pion'', constituted from mass-degenerate valence-partons with $s$-quark current-masses.

Returning now to Table~\ref{obsuds}B, we have not yet generalised the RL-truncation Faddeev equation, Fig.\,\ref{FEimage}, to the case of systems with non-degenerate valence-quark flavours.  This is formally straightforward.  However, as the study of mesons has shown, owing to moving singularities in the complex-$k^2$ domain sampled by the integration \cite{Maris:1997tm}, it can become difficult practically to obtain a reliable solution when the difference between the Euclidean constituent-quark masses of the valence-quarks involved becomes large; and, moreover, RL truncation becomes a poor approximation as one moves into the domain of heavy-light systems \cite{Bashir:2012fs}.
Consequently, the baryon mass predictions listed in the rightmost four columns of Table~\ref{obsuds}B were obtained using the equal spacing rule (ESR) described in connection with Eqs.\,\eqref{eqESR}.  Experience and usage indicates that these ESR values should reliably approximate the results that would be obtained directly from Fig.\,\eqref{FEimage}: at most, they may underestimate the RL-truncation Faddeev equation values by $1$-$2$\%.  This is smaller than the RL truncation's systematic error; an insight which \emph{a posteriori} justifies our decision not to become encumbered with the effort of solving the Faddeev equation for flavour-asymmetric systems.

\begin{table}[t]
\caption{\label{obsQmeson}
(A). Computed values for a range of properties of ground-state $S$-wave heavy-quarkonia, obtained using the quark-quark scattering kernel described in Sec.\,\ref{SeccalG} to specify the relevant gap- and Bethe-Salpeter-equations.  The interaction scale is stated in Eq.\,\eqref{varsigmaQ} and the current-quark masses listed in Eq.\,\eqref{currentQ} were chosen to reproduce the empirical values of $m_{\eta_c}$, $m_{\eta_b}$.  We used $\omega_Q=0.8$GeV; notably, a $\pm 10$\% change in this value has almost no perceptible effect on our results, e.g.\ $m_{\eta_c}=2.98(1)\,$GeV.
The rms relative-error per degree-of-freedom between calculation and experiment is 9\%.
(where reported, experimental values are inferred from Ref.\,\cite{Olive:2016xmw} -- $f_{\eta_c} = 0.238(12)$,
$f_{J/\Psi} = 0.294(5)$,
$f_{\Upsilon}=0.506(3)$;
and lattice-QCD (lQCD) results are drawn from Refs.\,\cite{Davies:2010ip, McNeile:2012qf, Donald:2012ga, Colquhoun:2014ica} -- $f_{\eta_c}=0.279(17)$, $f_{\eta_b}=0.472(4)$, $f_{J/\Psi}=0.286(4)$, $f_\Upsilon=0.459(22)$.)
(B)
Row~1 -- Computed RL truncation results for selected $B_c$ and $B_c^\ast$ meson properties;
Row~2 -- Estimates of these same quantities obtained using equal spacing rules, Eqs.\,\eqref{eqESR} -- \eqref{eqfKESR};
Row~3 -- Same rules used with experimental values of the relevant quarkonia properties;
Row~4 -- Results stated in Ref.\,\cite{Chiu:2007bc}.
An average of theory results yields $m_{B_c} = 6.336(2)$ \cite{Gomez-Rocha:2016cji}.
(All quantities listed in GeV.)
}
\begin{center}
\begin{tabular*}
{\hsize}
{l@{\extracolsep{0ptplus1fil}}
|l@{\extracolsep{0ptplus1fil}}
l@{\extracolsep{0ptplus1fil}}
l@{\extracolsep{0ptplus1fil}}
l@{\extracolsep{0ptplus1fil}}
l@{\extracolsep{0ptplus1fil}}
l@{\extracolsep{0ptplus1fil}}
l@{\extracolsep{0ptplus1fil}}
l@{\extracolsep{0ptplus1fil}}}\hline
(A)    & $m_{\eta_c}$ & $f_{\eta_c}$ & $m_{\eta_b}$ & $f_{\eta_b}$ & $m_{J/\psi}$ & $f_{J/\psi}$ & $m_\Upsilon$ & $f_\Upsilon$ \\
herein\; & \underline{$2.98$} & $0.28$ & \underline{$9.40$} & $0.57$ & $3.12$ & $0.30$ & $9.50$ & $0.54$  \\
expt. & $2.98$ & $0.24$ & $9.40$ & $-$ & $3.10$ & $0.29$ & $9.46$ & $0.51$ \\
lQCD &  & $0.28$ & & $0.47$ & & $0.29$ & & $0.46$ \\\hline
\end{tabular*}
\smallskip

\begin{tabular*}
{\hsize}
{l@{\extracolsep{0ptplus1fil}}
|l@{\extracolsep{0ptplus1fil}}
l@{\extracolsep{0ptplus1fil}}
l@{\extracolsep{0ptplus1fil}}
l@{\extracolsep{0ptplus1fil}}
l@{\extracolsep{0ptplus1fil}}
l@{\extracolsep{0ptplus1fil}}
l@{\extracolsep{0ptplus1fil}}
l@{\extracolsep{0ptplus1fil}}}\hline
(B)    & $m_{B_c}$ & $f_{B_c}$ & $m_{B_c^\ast}$ & $f_{B_c^\ast}$  \\
herein\;  & $6.39(1)$ & $0.43$ & $6.54(2)$ & $0.43$  \\
ESR\,herein\;  & $6.19(2)$ & $0.42$ & $6.31(2)$ & $0.42$  \\
ESR\,expt. \; & $6.19$ & $-$ & $6.28$ & $-$ \\
lQCD\; & $6.28(1)$ & $0.35$ & $6.32(1)$ & $-$ \\
expt. \; & $6.27$ & $-$ & $-$ & $-$ \\\hline
\end{tabular*}

\end{center}
\end{table}

\section{Heavy quark mesons}
\label{SecMesonHeavy}
As discussed at the end of Sec.\,\ref{SeccalG}, RL truncation should serve as a good approximation in the study of systems constituted from heavy quarks so long as the mass-scale in Eq.\,\eqref{varsigmaQ} is used.  In this case one must choose a value for $\omega$, \emph{i.e}.\ the range of the infrared piece of the interaction.  Considering again QCD's process-independent effective charge \cite{Binosi:2016nme}, a more realistic value of $\tilde\alpha_0/\tilde m_0$ is obtained by increasing $\omega$.  Following this guide, we incremented $\omega$, keeping $\varsigma=\varsigma_Q$ fixed, so as to optimise our description of the masses and decay constants of $S$-wave heavy-quarkonia.

With $\omega_Q = 0.8\,$GeV and the choices
\begin{equation}
\label{currentQ}
m_{c}^{\zeta_{19}} = 0.82\,{\rm GeV}\,,\; m_b^{\zeta_{19}} = 3.59\,{\rm GeV}\,,
\end{equation}
we obtain the results in Table~\ref{obsQmeson}.
The values in Eq.\,\eqref{currentQ} correspond to renormalisation-group-invariant masses $\hat m_{c}=1.61\,$GeV, $\hat m_b=7.16\,$GeV; one-loop-evolved masses at 2\,GeV of
\begin{equation}
\label{mQO}
m_{c}^{2\,{\rm GeV}}=1.22\,{\rm GeV}\,, \; m_{b}^{2\,{\rm GeV}}=5.41\,{\rm GeV};
\end{equation}
Euclidean constituent quark masses
\begin{equation}
M^E_c=1.32\,{\rm GeV}, \quad M^E_b=4.22\,{\rm GeV};
\end{equation}
and give $\hat m_c/\hat m_s=11$, $\hat m_b/\hat m_s=49$.
%
%
They are thus compatible with other contemporary estimates \cite{Olive:2016xmw}.

We have also computed the masses and decay constants of the $B_c$- and $B_c^\ast$-mesons in this RL truncation: our results are listed in Row~1 of Table\,\ref{obsQmeson}B.\footnote{The $B_c^\ast$-meson is only just accessible in a brute-force use of RL truncation, owing to the moving singularities described above \cite{Maris:1997tm}.}
For comparison, recall Eqs.\,\eqref{eqESR} -- \eqref{eqfKESR} and note that RL truncation predicts analogous linearities for $S$-wave heavy-heavy mesons \cite{Bhagwat:2004hn, Bhagwat:2006xi}.
(The correspondence does not extend to heavy-light systems \cite{Ivanov:1998ms}.)
This prediction of near-linearity leads to the estimates in Row~2 of Table~\ref{obsQmeson}B.  Notably, they are only $\sim 3$\% underestimates of the true RL results and the value of $m_{B_c}$ obtained in this way is within $1.2$\% of the current empirical value \cite{Olive:2016xmw}.
If one were interested in fine-tuning the ESR estimates, then the methods of potential non-relativistic QCD could be adopted \cite{Peset:2015vvi}.

\section{Triply-Heavy Baryons}
\label{SecQQQ}
\subsection{Mass}
\begin{figure}[!t]
\begin{center}
\includegraphics[clip,width=\linewidth]{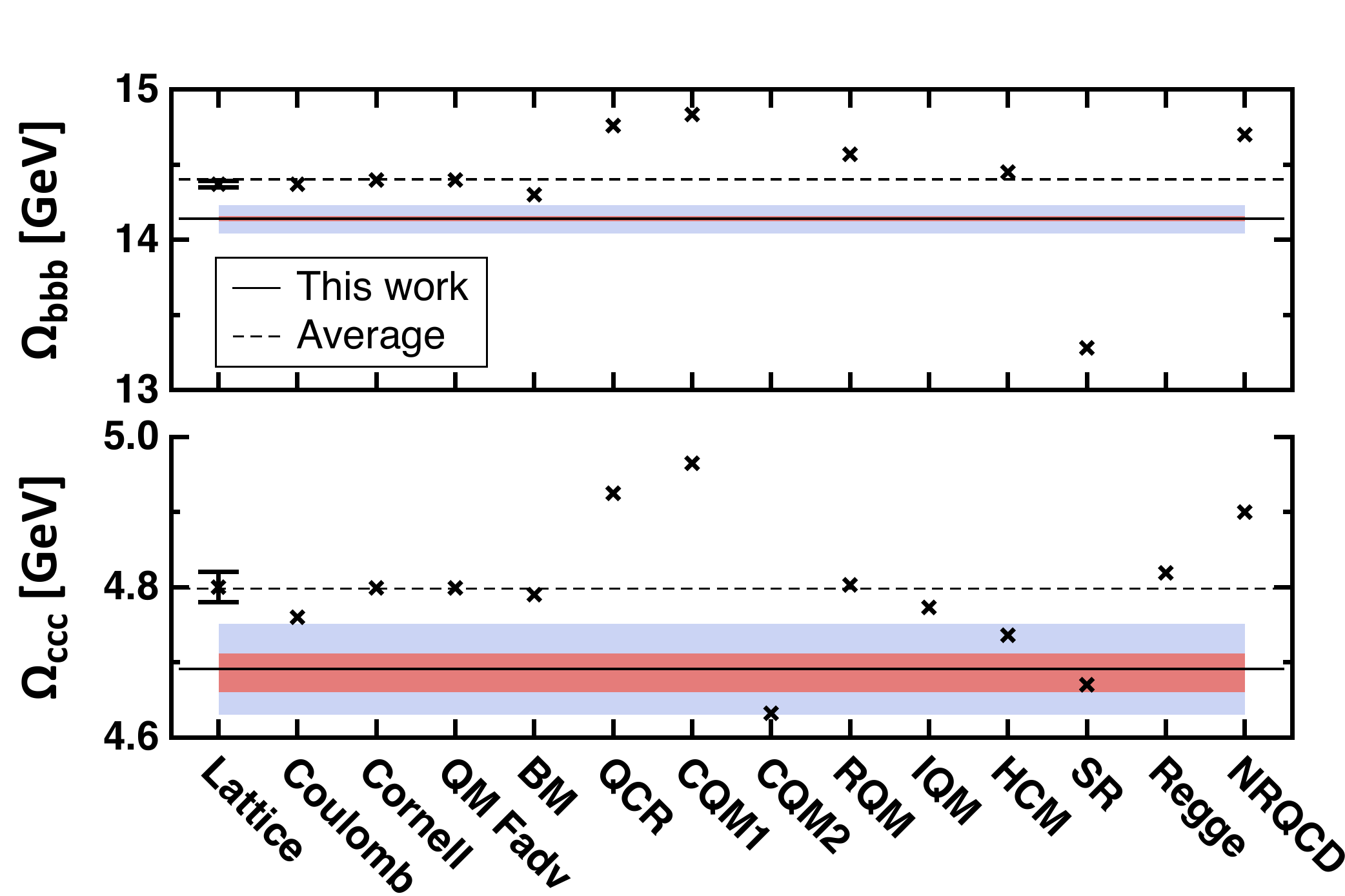}\vspace*{-6ex}

\leftline{\textbf{(A)}}

\medskip

\includegraphics[clip,width=\linewidth]{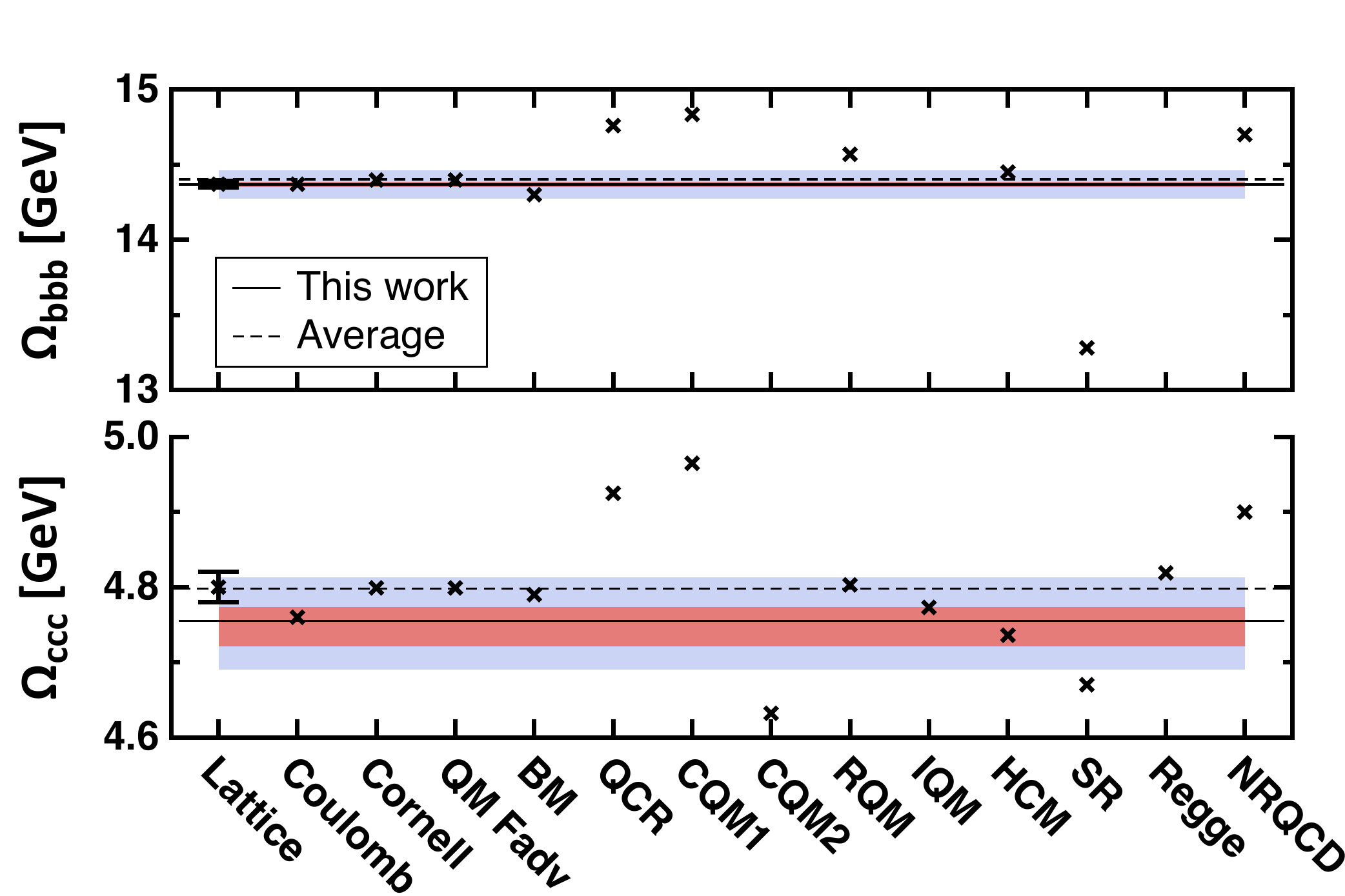}\vspace*{-6ex}

\leftline{\textbf{(B)}}

\end{center}
\caption{\label{CompOmega1}
\emph{Upper panel} (A) -- Our predictions for the $\Omega_{ccc}$ and $\Omega_{bbb}$ masses: darker (red) band, impact of $\omega_Q = 0.8 \pm 0.08$ ($\pm 10$\%); and lighter (blue) band, $\pm 10$\% change to $\Lambda_{\rm QCD}$ in Eq.\,\eqref{LambdaQCD}.  The masses increase with either increase in scale.
For comparison, we also indicate results obtained using other approaches, left-to-right:
Refs.\,\cite{Brown:2014ena}, 
\cite{Jia:2006gw, 
SilvestreBrac:1996bg, 
Hasenfratz:1980ka, 
Bjorken:1985ei}, 
\cite{Roberts:2007ni}, 
\cite{Vijande:2004at, 
Martynenko:2007je, 
Migura:2006ep, 
Patel:2008mv, 
Zhang:2009re, 
Guo:2008he}, 
\cite{LlanesEstrada:2011kc}.  
\emph{Lower panel} (B)  -- Analogous comparison, prepared using the current-quark masses in Eq.\,\eqref{currentQR}, which yield the baryon masses in the second rows of Table~\ref{obsQbaryon}(A,B).
}
\end{figure}

Using the quark-quark scattering kernel constrained via the study of heavy-heavy mesons, we solved the Faddeev equation depicted in Fig.\,\ref{FEimage} to obtain the masses and Faddeev amplitudes of $J^P=\tfrac{3}{2}^+$ $ccc$ and $bbb$ baryons: our results are compared with a raft of other estimates in Fig.\,\ref{CompOmega1} and listed in Table~\ref{obsQbaryon}.

The comparison in Fig.\,\ref{CompOmega1}A reveals that, using the current-quark masses in Eq.\,\eqref{currentQ}, our results for the $\Omega_{ccc}$ and $\Omega_{bbb}$ masses are low compared with a theory average.  We therefore repeated all relevant calculations using current-quark masses inflated by 2\%:
\begin{equation}
\label{currentQR} 
m_{c}^{\zeta_{19}} = 0.83\,{\rm GeV}\,,\; m_b^{\zeta_{19}} = 3.66\,{\rm GeV}\,.
\end{equation}
This yields the masses listed in the second rows of Table~\ref{obsQbaryon}(A,B), from which we produced Fig.\,\ref{CompOmega1}B: on average, the values are increased by 1.5\%.  Evidently, this small increase in current-quark mass is sufficient to remedy the mass-deficit with respect to the theory average.

It is worth noting here that the current-quark masses in Eq.\,\eqref{currentQR} correspond
to renormalisation-group-invariant masses $\hat m_{c}=1.64\,$GeV, $\hat m_b=7.30\,$GeV; one-loop-evolved masses at 2\,GeV of
\begin{equation}
\label{mQOO}
m_{c}^{2\,{\rm GeV}}=1.24\,{\rm GeV}\,, \; m_{b}^{2\,{\rm GeV}}=5.52\,{\rm GeV};
\end{equation}
Euclidean constituent quark masses
\begin{equation}
M^E_c=1.35\,{\rm GeV}, \quad M^E_b=4.28\,{\rm GeV};
\end{equation}
and give $\hat m_c/\hat m_s=11$, $\hat m_b/\hat m_s=50$.
Computed with these current-quark masses, the values of heavy-heavy meson observables reported in Table~\ref{obsQmeson} increase by 1.6\%, on average.  Such changes are within the margin of error for RL truncation.

For the reasons detailed in closing Sec.\,\ref{SecMeson}, we do not directly solve the Faddeev equation in Fig.\,\ref{FEimage} for $ccb$ and $cbb$ baryons, but instead report masses obtained using equal spacing rules analogous to Eqs.\,\eqref{eqESR}: at worst, these values might underestimate the true RL results by $\sim 3$\%, an amount subsumed in the reported error.

Tables~\ref{obsQbaryon}(C,D) reexpress all computed results in Tables~\ref{obsQbaryon}(A,B) as a multiple of the $\Omega_{ccc}$ mass computed within the same framework/setup.  Evidently, so far as these systems are concerned, the gross features of the spectrum are fixed once the overall mass-scale is set; and hence results for level splittings are very sensitive to fine details of the interaction.  Such precision is beyond the scope of RL truncation; and might also be a challenge to other contemporary nonpeturbative approaches to strong QCD.

\begin{table}[t]
\caption{\label{obsQbaryon}
Baryon masses obtained by solving the Faddeev equation in Fig.\,\ref{FEimage} using the quark-quark scattering kernel described in Sec.\,\ref{SecMesonHeavy}:
(A) -- $J=\tfrac{3}{2}^+$ ground state; and (B) -- $J=\tfrac{3}{2}^+$ first positive-parity excitation.  Two sets of current-quark masses are used: Eqs.\,\eqref{currentQ}, \eqref{currentQR}.
Our results in the two rightmost columns of panels (A) and (B) were obtained using equal spacing rules analogous to Eqs.\,\eqref{eqESR}; and the masses are listed in GeV.
Panels (C) and (D) reexpress the results in (A) and (B) in terms of the like-computed $\Omega_{ccc}$ mass.
}
\begin{center}
\begin{tabular*}
{\hsize}
{l@{\extracolsep{0ptplus1fil}}
|l@{\extracolsep{0ptplus1fil}}
l@{\extracolsep{0ptplus1fil}}
|l@{\extracolsep{0ptplus1fil}}
l@{\extracolsep{0ptplus1fil}}}\hline
(A)       & $\Omega_{ccc}$ & $\Omega_{bbb}$ & $\Omega_{ccb}$ & $\Omega_{cbb}$    \\
herein -- \eqref{currentQ}\; & $4.69(6)$ & $14.14(10)$\; & $7.84(12)$ & $10.99(12)$  \\
herein -- \eqref{currentQR}\; & $4.76(7)$  & $14.37(10)$ & $7.96(12)$ & $11.17(12)$ \\
lQCD \cite{Brown:2014ena} & $4.80(2)$ & $14.37(2)$ & $8.01(2)$ & $11.20(2)$  \\\hline
\end{tabular*}
\medskip

\begin{tabular*}
{\hsize}
{l@{\extracolsep{0ptplus1fil}}
|l@{\extracolsep{0ptplus1fil}}
l@{\extracolsep{0ptplus1fil}}
|l@{\extracolsep{0ptplus1fil}}
l@{\extracolsep{0ptplus1fil}}}\hline
 (B)     & $\Omega_{ccc}^\prime$ & $\Omega_{bbb}^\prime$ & $\Omega_{ccb}^\prime$ & $\Omega_{cbb}^\prime$   \\
herein -- \eqref{currentQ}\; & $5.08(8)$ & $14.74(12)\;$ & $8.30(14)$ & $11.52(14)$   \\
herein -- \eqref{currentQR}\; & $5.15(8)$ & $14.98(12)$ & $8.47(14)$ & $11.76(14)$   \\
%
\hline\hline
\end{tabular*}
\medskip

\begin{tabular*}
{\hsize}
{l@{\extracolsep{0ptplus1fil}}
|l@{\extracolsep{0ptplus1fil}}
|l@{\extracolsep{0ptplus1fil}}
l@{\extracolsep{0ptplus1fil}}
l@{\extracolsep{0ptplus1fil}}}\hline\hline
(C)       & $\Omega_{ccc}$ & $\Omega_{bbb}$ & $\Omega_{ccb}$ & $\Omega_{cbb}$    \\
herein -- \eqref{currentQ}\; & reference\; & $3.01(5)$\; & $1.67(3)$ & $2.34(4)$  \\
herein -- \eqref{currentQR}\; & \rule{1.5em}{0ex}''  & $3.02(5)$ & $1.67(4)$ & $2.35(4)$ \\
lQCD \cite{Brown:2014ena} & \rule{1.5em}{0ex}'' & $2.99(1)$ & $1.67(1)$ & $2.33(1)$  \\\hline
\end{tabular*}

\medskip

\begin{tabular*}
{\hsize}
{l@{\extracolsep{0ptplus1fil}}
|l@{\extracolsep{0ptplus1fil}}
l@{\extracolsep{0ptplus1fil}}
l@{\extracolsep{0ptplus1fil}}
l@{\extracolsep{0ptplus1fil}}}\hline
 (D)     & $\Omega_{ccc}^\prime$ & $\Omega_{bbb}^\prime$ & $\Omega_{ccb}^\prime$ & $\Omega_{cbb}^\prime$   \\
herein -- \eqref{currentQ}\; & $1.08(2)$ & $3.14(5)\;$ & $1.77(4)$ & $2.46(4)$   \\
herein -- \eqref{currentQR}\; & $1.08(2)$ & $3.15(5)$ & $1.78(4)$ & $2.47(5)$   \\
%
\hline
\end{tabular*}
\end{center}
\end{table}

The success of Eqs.\,\eqref{eqESR} motivates us to define a constituent-quark passive-mass via
\begin{equation}
\label{EqMfP}
M_f^P = \tfrac{1}{3} m_{\Omega_{fff}}\,,
\end{equation}
with the computed values (in GeV):
\begin{equation}
\label{EqMfP2}
{\rm baryon:}\quad
\begin{array}{l|cccc}
f & u=d & s & c & b \\\hline
M_f^P & 0.40 & 0.56 & 1.56 & 4.71
\end{array}\,.
\end{equation}
The analogous quantity defined via ground-state vector-meson masses takes very similar values (in GeV):
%
\begin{equation}
\label{EqMfP2meson}
{\rm meson:~}\quad
\begin{array}{l|cccc}
f & u=d & s & c & b \\\hline
M_f^P & 0.38 & 0.55 & 1.56 & 4.75
\end{array}\,.
\end{equation}
For light quarks in RL truncation, $M_q^P$ is close to the Euclidean constituent-quark mass, but $M_Q^P$ is better matched with $M_Q(k\simeq 0)$, \emph{i.e}.\ the appropriate heavy-quark mass-function evaluated near the origin.  In DB-truncations, the latter is true for all quark flavours because $M_q(0)$ and $M^E_q$ are more nearly equal for sound physical reasons that are understood \cite{Chang:2009zb, Chang:2010hb, Chang:2011ei, Binosi:2014aea, Binosi:2016wcx, Binosi:2016nme}.

We now capitalise on Eqs.\,\eqref{EqMfP}, \eqref{EqMfP2}, using these constituent-quark passive-masses to obtain mass estimates for the members of the symmetric-$\mathbf{20}$ of $SU_c(4)$ and $SU_b(4)$ that we have not computed directly.  The results are reported in Table~\ref{Symmetric20}: evidently, the equal spacing rules also provide a reasonable guide for these systems.

\begin{table}[t]
\caption{
\label{Symmetric20}
Baryon mass estimates obtained using the equal spacing rule constituent-quark passive-masses in Eqs.\,\eqref{EqMfP}, \eqref{EqMfP2} compared with values from a contemporary simulation of lattice-QCD \cite{Brown:2014ena}:
(A) -- $c$-quark systems; (B) -- $b$-quarks.
(All quantities listed in units of $m_{\Omega_{ccc}}$ as obtained in the cited work and reported in Table~\ref{obsQbaryon}.)
}
\begin{center}
\begin{tabular*}
{\hsize}
{l@{\extracolsep{0ptplus1fil}}
|l@{\extracolsep{0ptplus1fil}}
l@{\extracolsep{0ptplus1fil}}
l@{\extracolsep{0ptplus1fil}}
l@{\extracolsep{0ptplus1fil}}}\hline
(A)       & $\Sigma_{uuc}$ & $\Xi_{ucc}$ & $\Omega_{ssc}$ & $\Omega_{scc}$    \\
herein -- \eqref{EqMfP2}\; & $0.51(1)$ & $0.75(1)$\; & $0.57(1)$ & $0.79(1)$   \\
lQCD \cite{Brown:2014ena} & $0.52(1)$ & $0.75(1)$ & $0.56(2)$ & $0.78(1)$    \\\hline
\end{tabular*}
\medskip

\begin{tabular*}
{\hsize}
{l@{\extracolsep{0ptplus1fil}}
|l@{\extracolsep{0ptplus1fil}}
l@{\extracolsep{0ptplus1fil}}
l@{\extracolsep{0ptplus1fil}}
l@{\extracolsep{0ptplus1fil}}}\hline
 (B)      & $\Sigma_{uub}$ & $\Xi_{ubb}$ & $\Omega_{ssb}$ & $\Omega_{sbb}$    \\
herein -- \eqref{EqMfP2}\; & $1.18(2)$ & $2.10(3)\;$ & $1.24(2)$ & $2.13(3)$    \\
lQCD \cite{Brown:2014ena} & $1.22(1)$ & $2.11(1)$ & $1.26(1)$ & $2.14(1)$   \\\hline
%
\hline
\end{tabular*}

\end{center}
\end{table}

\begin{figure*}[!t]
\begin{center}
\begin{tabular}{lr}
\includegraphics[clip,width=0.47\linewidth]{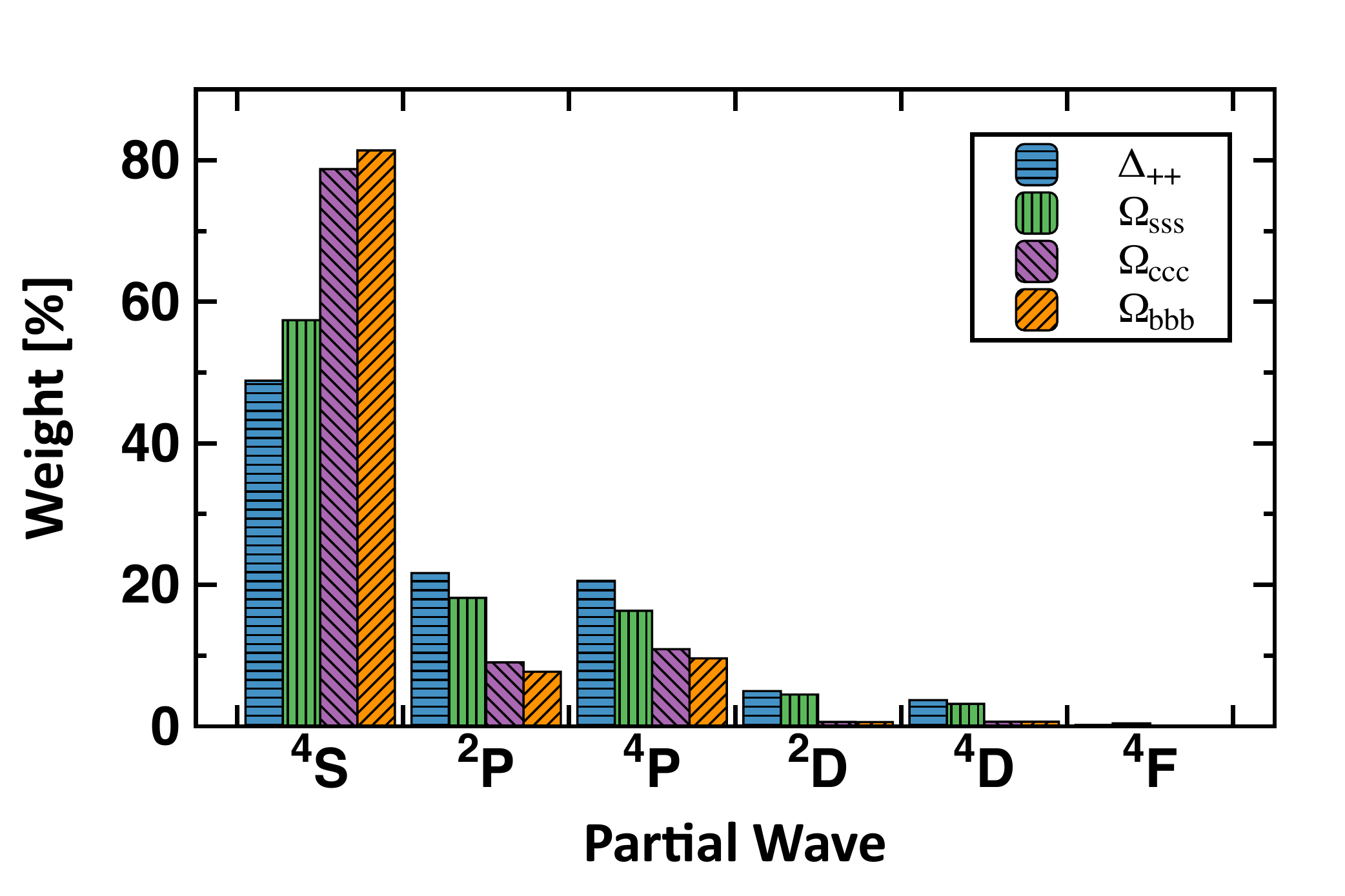}\hspace*{2ex } &
\includegraphics[clip,width=0.47\linewidth]{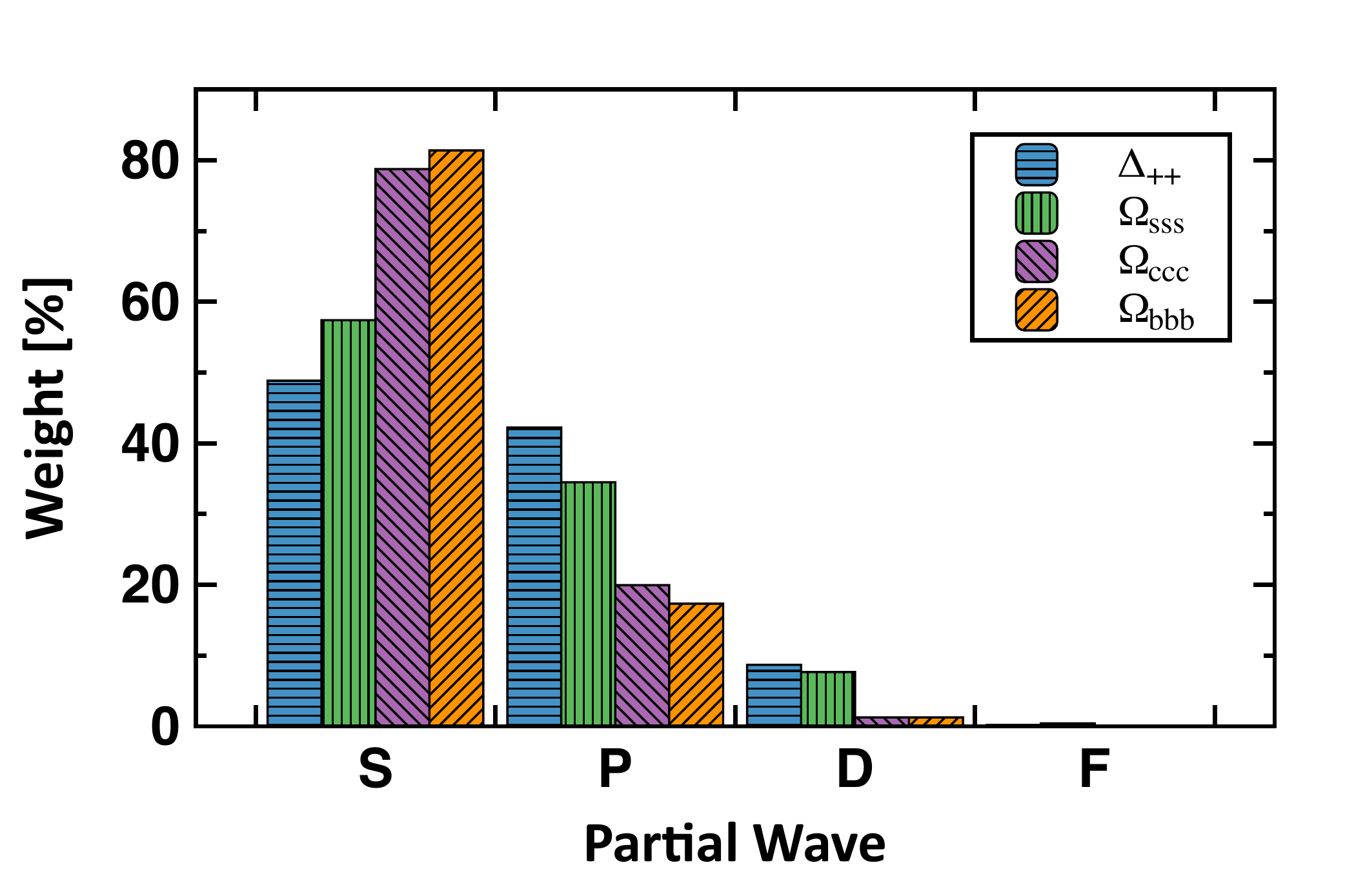}\vspace*{-0ex}
\end{tabular}
\begin{tabular}{lr}
\includegraphics[clip,width=0.47\linewidth]{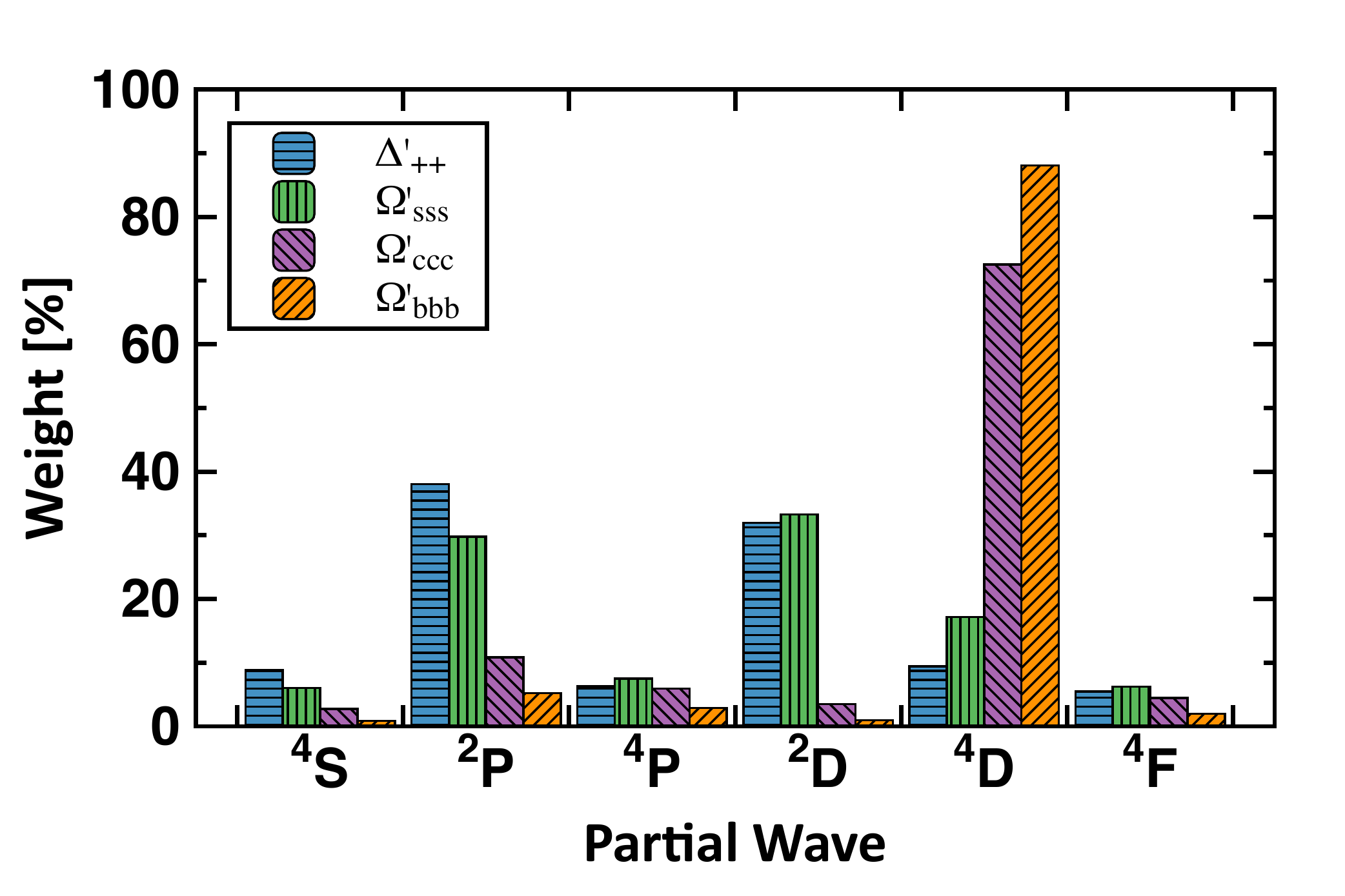}\hspace*{2ex } &
\includegraphics[clip,width=0.47\linewidth]{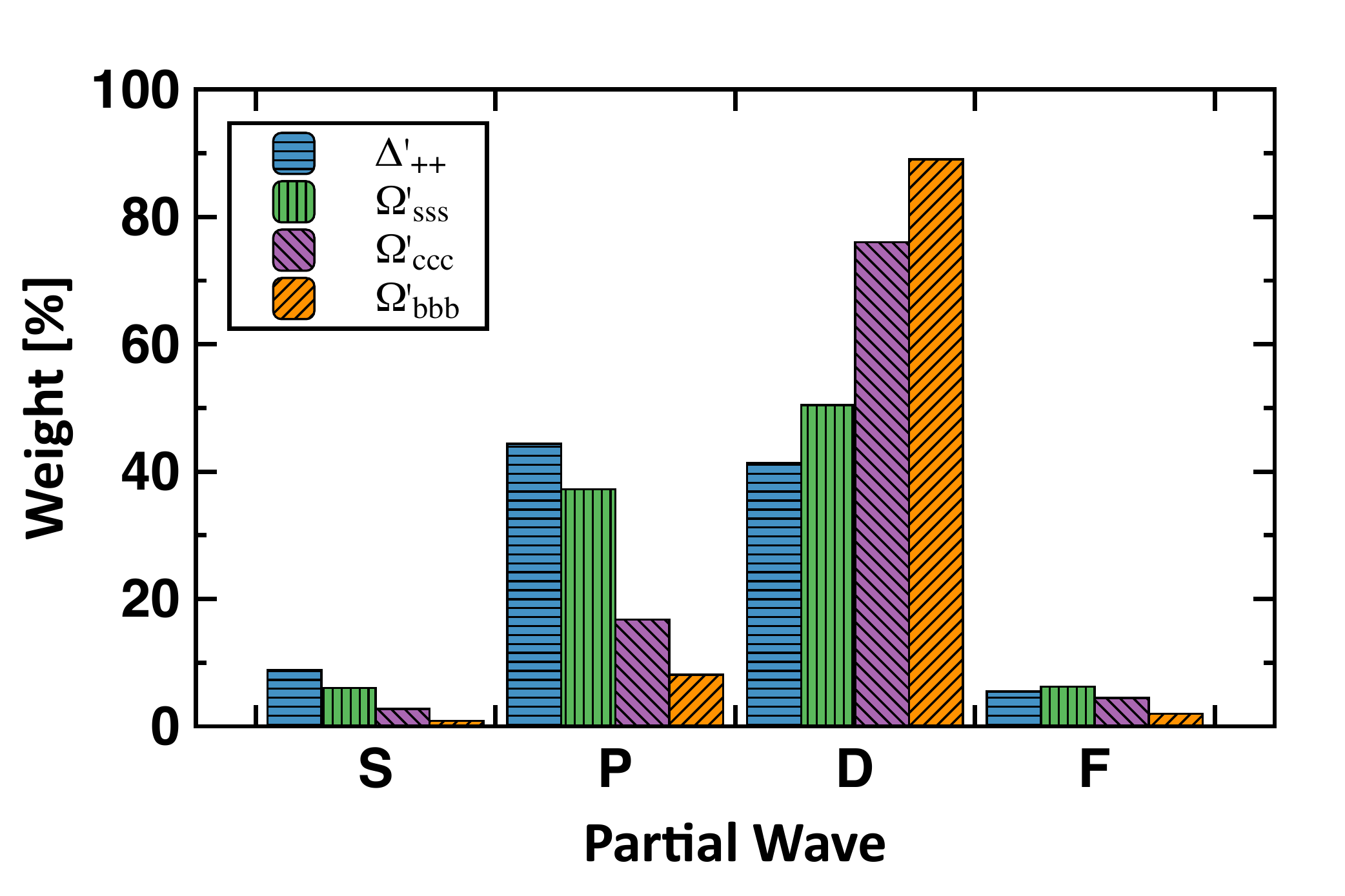}\vspace*{-1ex}
\end{tabular}
\end{center}
\caption{\label{PWDs}
Baryon rest-frame three-body angular-momentum fractions, Eq.\,\eqref{Wproject}:
\emph{Upper panels} -- ground state $J=3/2^+$: \emph{left}, complete decomposition; and \emph{right}, gross features.
\emph{Lower panels} -- same for the first positive-parity excitation of each state. }
\end{figure*}

The results in Table~\ref{obsQbaryon} predict that the energy cost of a baryon's first positive-parity excitation rises slowly with increasing current-quark mass (GeV):
\begin{equation}
\label{radial}
\begin{array}{rcl}
m_{\Delta^\prime} - m_{\Delta} &=& 0.25\,,\;\\
m_{\Omega^\prime} - m_{\Omega} &=& 0.29\,,\;\\
m_{\Omega_{ccc}^\prime} - m_{\Omega_{ccc}} &=& 0.39(1)\,,\;\\
m_{\Omega_{bbb}^\prime} - m_{\Omega_{bbb}} &=& 0.60(16)\,.
\end{array}
\end{equation}
There is insufficient empirical information to test this prediction in baryons.  The only comparison one can currently draw is with the evolution of the analogous splitting between degenerate-flavour vector mesons, for which the pattern is a little different, \emph{viz}.\ the splitting falls slowly on the empirically accessible domain.  Notably, $m_{\Upsilon(2S)}-m_{\Upsilon(1S)} = 0.56\,$GeV, which is similar to the value of $m_{\Omega_{bbb}^\prime} - m_{\Omega_{bbb}}$ in Eq.\,\eqref{radial}.

\subsection{Rest-frame orbital angular momentum}
Beginning with the baryon Faddeev amplitude in Eq.\,\eqref{FaddeevAmp}, one can construct the Faddeev wave function by attaching the dressed-quark legs:
\begin{align}
\nonumber
 {\chi}& \!\,^{\alpha_1\alpha_2\alpha_3,\delta}_{\mu}(p_1,p_2,p_3) = S_{\alpha_1\alpha_2'}(p_1)  \\
	& \times S_{\alpha_2\alpha_2'}(p_2) S_{\alpha_3\alpha_3'}(p_3)
{\Psi}^{\alpha_1^\prime \alpha_2^\prime \alpha_3^\prime,\delta }_{\mu}(p_1,p_2,p_3)\,, \label{FaddeevWaveF}
\end{align}
and thereby arrive at that quantity which can directly be compared with the system's Schr\"odinger wave function when a non-relativistic limit is valid.  Both the Faddeev amplitude and wave function are Poincar\'e covariant, \emph{i.e}.\ they are qualitatively identical in all reference frames.  Consequently, each of the scalar functions that appears is frame-independent, but the frame chosen determines just how the elements should be combined.  In consequence, the manner by which the dressed-quarks' spin, $S$, and orbital angular momentum, $L$, add to form $J^P=3/2^+$ is frame-dependent: $L$, $S$ are not independently Poincar\'e invariant.

One can enable comparisons with typical nonrelativistic treatments by working with $\chi$ evaluated in the bound-state's rest frame.  It may then be decomposed into a sum of six terms, each one representing a distinct $L$-$S$ coupling from the following list:
\begin{equation}
^{2 S+1}\!L_{J=\tfrac{3}{2}}  \to  (\,^4\!S_{\tfrac{3}{2}}\,,\;
^2\!P_{\tfrac{3}{2}}\,,\; ^4\!P_{\tfrac{3}{2}}\,,\; ^2\!D_{\tfrac{3}{2}}\,,\; ^4\!D_{\tfrac{3}{2}}\,,\; ^4\!F_{\tfrac{3}{2}})\,.
\end{equation}
(We shall subsequently omit the subscript $J=\tfrac{3}{2}$.)  In order to achieve this decomposition, we first use the structures explained in Appendix~\ref{FAmplitude} to write
\begin{align}
{\chi}^{\alpha_1\alpha_2\alpha_3,\delta}_{\mu}(p_1,p_2,p_3)
& =\sum_{n=1}^{128} {\mathpzc x}_n({\mathpzc z})\; [\mathsf{X}_n]_\mu^{\alpha_1\alpha_2\alpha_3,\delta}(\hat{{\mathpzc z}})\,;
\label{EqChi}
\end{align}
and then define the following array of rest-frame angular momentum strengths:
{\allowdisplaybreaks
\begin{subequations}
\label{Wproject}
\begin{align}
\,^4{\mathbb S} & = {\mathbb N}^{-1} \sum_{n \in\, ^4\!S}
    \int \frac{d^4 p}{(2\pi)^4}\frac{d^4 q}{(2\pi)^4} |{\mathpzc x}_n(p,q,P)|^2 \,,\\
\,^2{\mathbb P} & = {\mathbb N}^{-1} \sum_{n \in\, ^2\!P}
    \int \frac{d^4 p}{(2\pi)^4}\frac{d^4 q}{(2\pi)^4} |{\mathpzc x}_n(p,q,P)|^2 \,,
\end{align}
etc., where each sum runs only over those basis vectors which describe the specified rest-frame $L$-$S$ coupling, and
\begin{align}
{\mathbb N} & = \sum_{n=1}^{128} \int \frac{d^4 p}{(2\pi)^4}\frac{d^4 q}{(2\pi)^4} |{\mathpzc x}_n(p,q,P)|^2\,.
\end{align}
\end{subequations}}
\hspace*{-0.5\parindent}(This procedure is a three-body analogy of that employed elsewhere \cite{Chen:2017pse} to display the angular momentum content of the four lightest $(I=1/2,J^P=1/2^\pm)$ baryon isospin doublets.)

Computed using Eqs.\,\eqref{Wproject}, our values for the rest-frame three-body angular-momentum fractions in the $3/2^+$ baryon ground-states and their first positive-parity excitations are depicted in Fig.\,\ref{PWDs}: unsurprisingly in a Poincar\'e-covariant treatment, every possible partial wave is present, although the $F$-wave is small.
Notably, although the $S$-wave component is largest in each ground-state light-quark system, they also possess large net $P$-wave fractions, part of which enhancement owes to the presence of significantly more $P$-wave basis elements in Eq.\,\eqref{EqChi} than $S$-wave contributions.
%
Notwithstanding this, the $P$-wave component decreases steadily with increasing current-quark mass so that heavy-quark ground states are predominantly $S$-wave in nature.  Analogous evolution is seen with the $S$- and $D$-wave fractions in vector mesons constituted from degenerate valence partons \cite{Gao:2014bca}.

The structure of the positive-parity excitations is quite different (Fig.\,\ref{PWDs}, lower panels): $P$-waves dominate in the lightest system, but $D$-waves grow in strength with increasing current-quark mass, dominating in the $\Omega_{sss}^\prime$ and heavier systems.
Here, as with other observables, $s$-quark systems define a ``boundary'':  they are neither truly light nor heavy, and therefore provide a benchmark for comparisons between the impacts of strong and weak mass generation \cite{Ding:2015rkn}.

In connection with light-quark baryons, these observations are semi-quantitatively in agreement with those in Ref.\,\cite{Eichmann:2016hgl} even though a different measure of angular-momentum content was employed.  That study did not consider heavy baryons.

\subsection{Character of the first positive-parity excitations}
%
%
At this point it is worth answering a question; \emph{viz}.\ in the Poincar\'e-covariant treatment of these bound-states, may one interpret the first positive-parity excitation in each channel as a radial excitation or are these states more properly understood as even-parity $(L,S)$ ($D$-wave) excitations?  Comparing the two rightmost panels in Fig.\,\ref{PWDs}, it is evident that the $P$-wave fractions in the ground- and excited-states are quite similar, whereas the $S$-waves are much diminished and the $D$-waves greatly enhanced in the excited states.

\begin{figure}[!t]
\begin{center}
\includegraphics[clip,width=\linewidth]{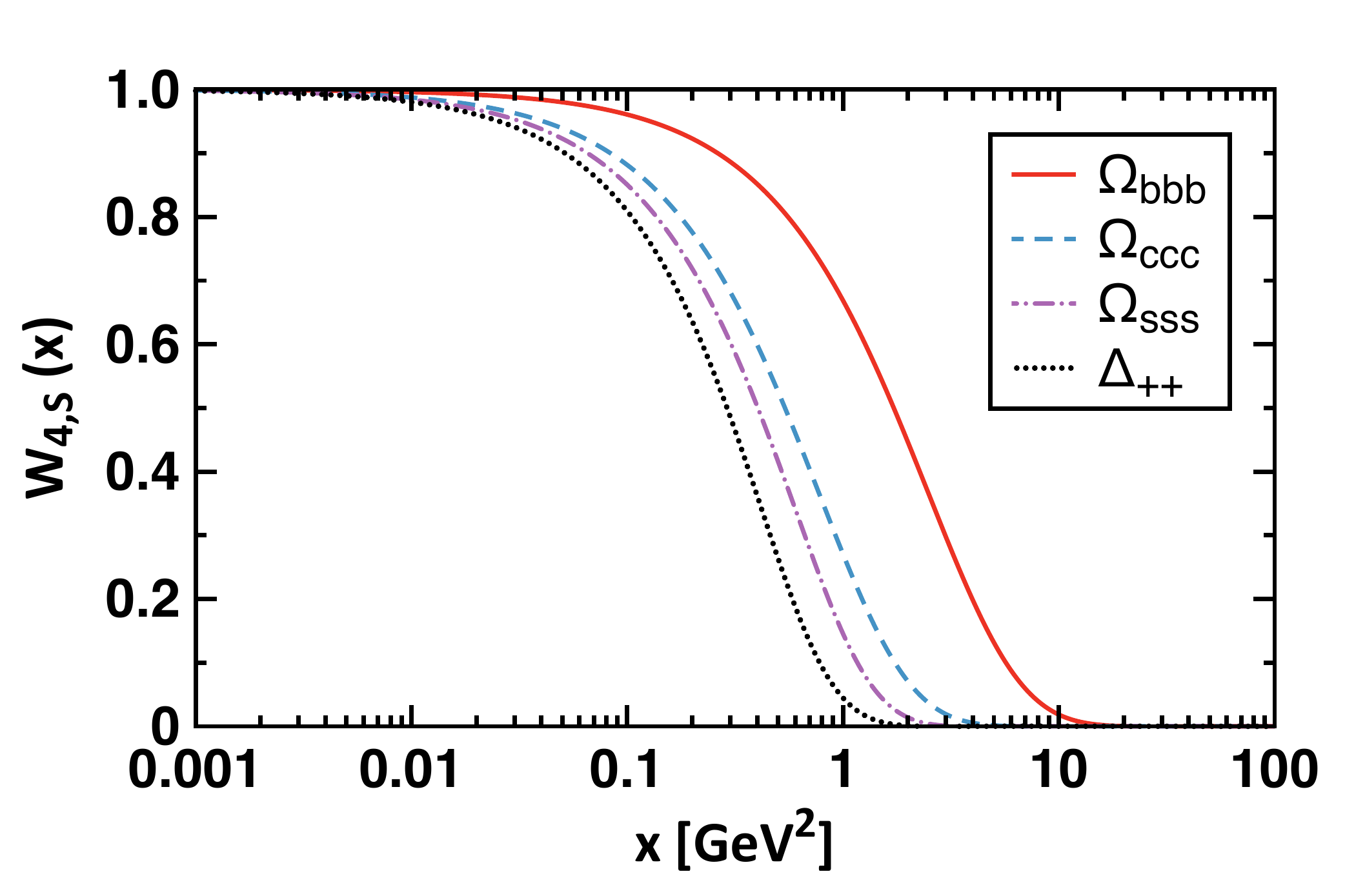}\vspace*{-6ex}
\end{center}
\caption{\label{FigWaveFunction} $S$-wave component of the Faddeev wave function for each $J^P=3/2^+$ baryon, Eq.\,\eqref{EqWFS}, normalised to unity at $x=0$.
}
\end{figure}

Additional information can be gleaned by studying the wave function of each system.  Therefore, in Fig.\,\ref{FigWaveFunction} we depict the $S$-wave component of each ground-state $3/2^+$ baryon, defined as follows:
\begin{align}
\label{EqWFS}
W_{4,S}(x) & = \sum_{n \in \! \, ^4\!S}  |{\mathpzc x}_n(x,0,0,0,0)|^2\,.
\end{align}
(Recall Eq.\,\eqref{EqPsivarsigma}.)  Not one of the functions drawn possesses a node on $x>0$.  Furthermore, we have checked each one of the 128 components of the ground-state baryon rest-frame-projected Faddeev wave functions and found that none possesses a node.

\begin{figure}[!t]
\begin{center}
\smallskip

\mbox{$\Delta^\prime$}
\includegraphics[clip,width=\linewidth]{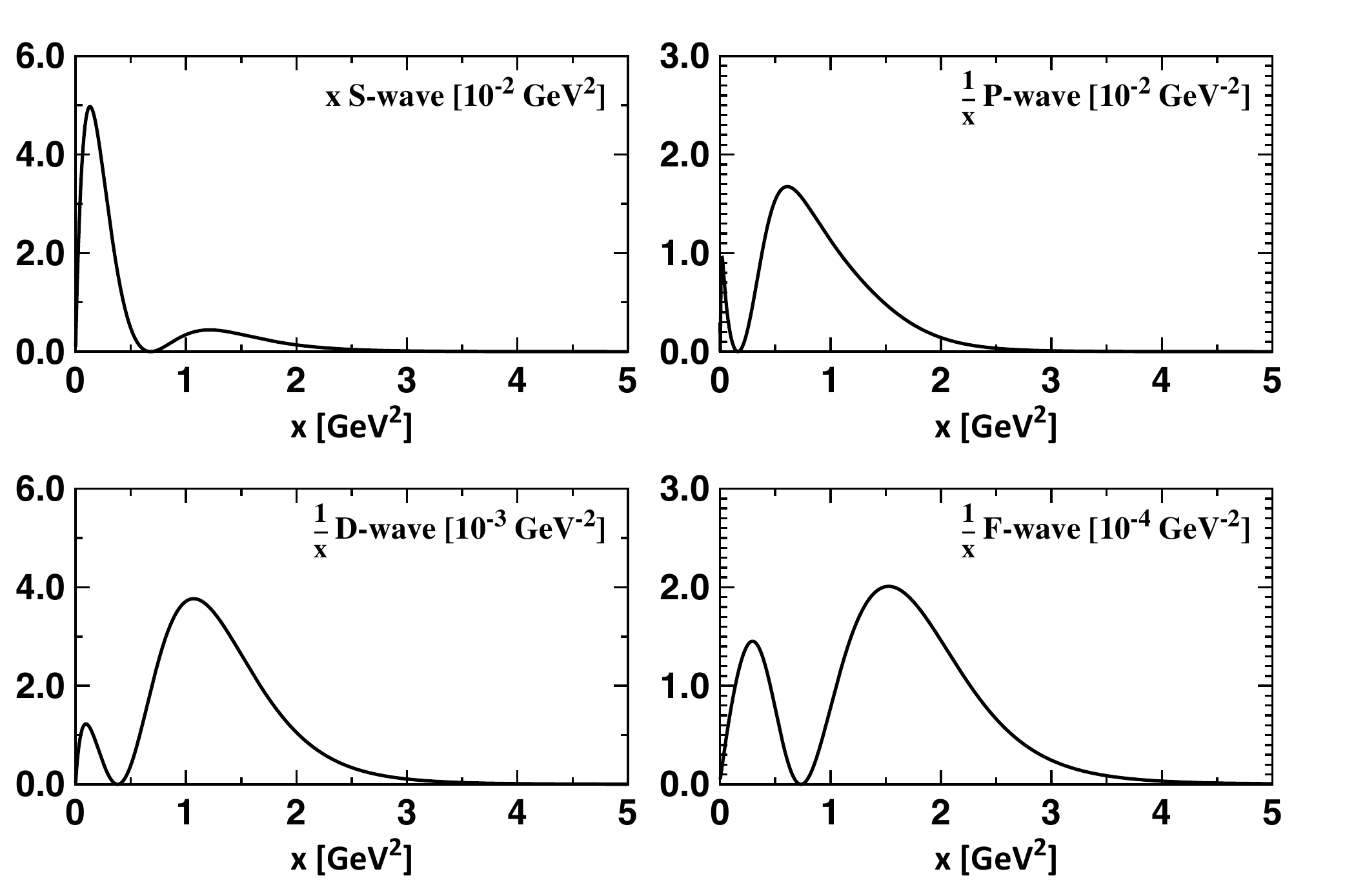} 

\mbox{$\Omega^\prime_{ccc}$}
\includegraphics[clip,width=\linewidth]{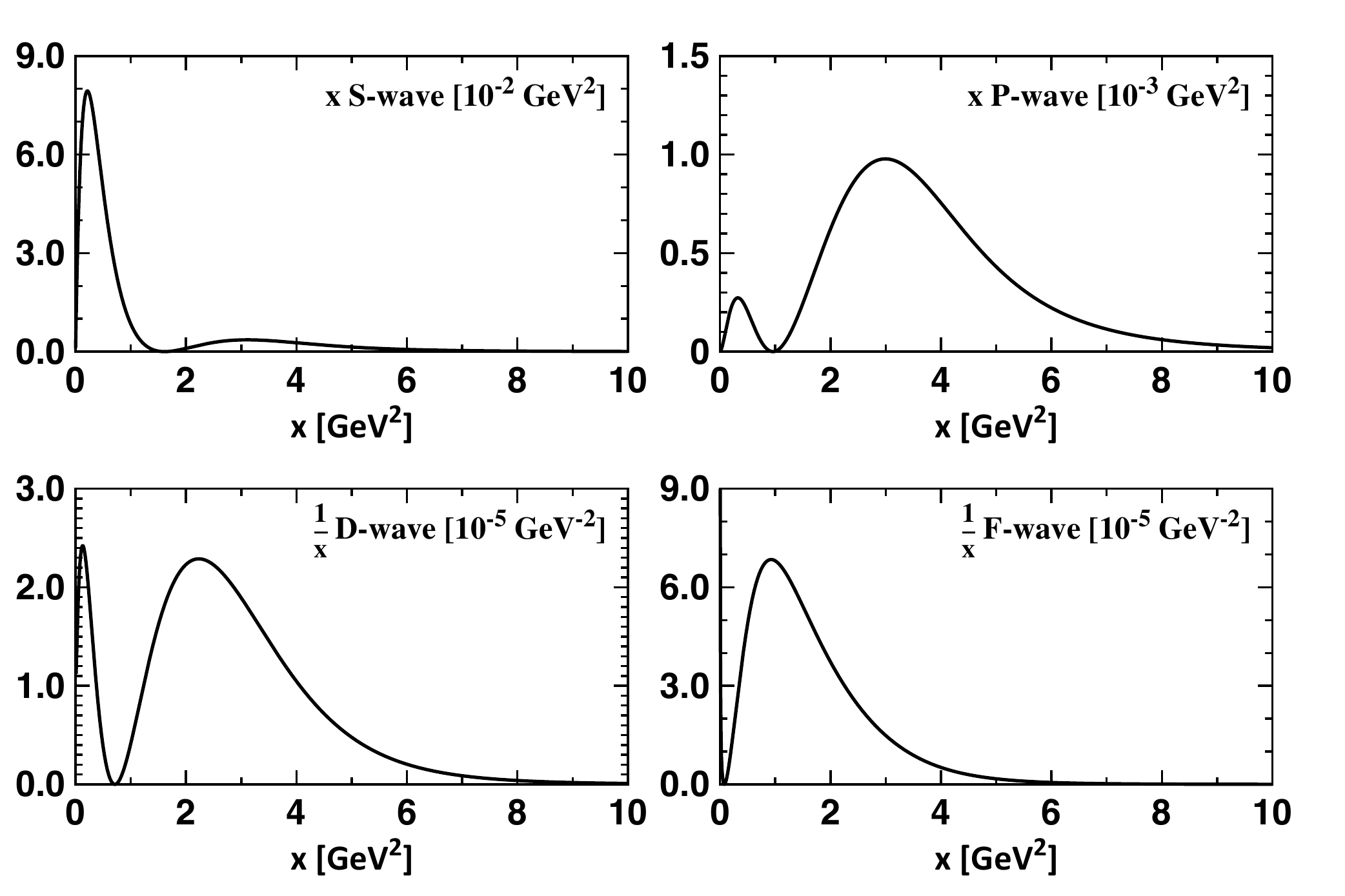}
\end{center}
\caption{\label{FigWaveFunctionE}
\emph{Upper four panels} -- $\Delta_{uuu}^\prime$; and \emph{lower four panels} -- $\Omega_{ccc}^\prime$.
Working with the rest-frame-projected Faddeev wave functions of the bound-states indicated, evaluated at ${\mathpzc z}=(x,0,0,0,0)$, each panel depicts $|\chi_n({\mathpzc z})|^2$ for the largest-magnitude three-dressed-quark angular-momentum component within each partial wave that possesses a zero.  (The $x$-weightings are chosen individually for visual effect.)
The Faddeev wave function is normalised such that the dominant $S$-wave component is unity at ${\mathpzc z}=(0,0,0,0,0)$. }
\end{figure}

The curves in Fig.\,\ref{FigWaveFunction} should be compared with those derived from the wave functions of the first positive-parity excitations.  For each of the $uuu$, $sss$, $ccc$, $bbb$ positive-parity excitations, we have checked all 128 components of the associated rest-frame-projected Faddeev wave function and learnt that with every component there is always at least one kinematic configuration for which it exhibits a single node.  Choosing the kinematic configuration ${\mathpzc z}=(x,0,0,0,0)$, then Fig.\,\ref{FigWaveFunctionE} depicts that individual function of largest magnitude within each partial wave projection which possesses a node.
There are no configurations ${\mathpzc z}$ for which any one the 128 amplitudes possesses more than one node on $x>0$.
%
%
%
Drawing upon experience with quantum mechanics and studies of excited-state mesons using the Bethe-Salpeter equation \cite{Holl:2004fr, Qin:2011xq, Rojas:2014aka, Li:2016dzv, Li:2016mah},  these features are indicative of a first radial excitation.

Combining this information, a picture of the first positive-parity excitations emerges.  In our Poincar\'e-covariant treatment in RL truncation, they are neither purely radial nor purely $D$-wave excitations; but, instead, they are a more intricate combination of both.
It is reported \cite{CChenPrivate} that a similar picture emerges when these systems are analysed using the quark-diquark approximation to the Faddeev equation described in Ref.\,\cite{Chen:2017pse}.

In nonrelativistic quark models \cite{Roberts:2007ni}, one typically finds that the first positive-parity excitation in the triply-heavy sector is primarily a radial excitation, with small $D$-wave admixtures.  However, that weighting is reversed in the second positive-parity excitation, the splitting between these states is small, and it decreases with increasing constituent quark mass.  It is possible, therefore, that the ordering is decided by fine details of the interaction.  Notwithstanding that, regarding the first postive-parity excitation, at this point there is an apparent quantitative contrast with our results; and whether it is real and significant may only be judged after, \emph{e.g}.\ predictions for decays and transitions are also compared.


\subsection{Spacetime extent}
We return now to Fig.\,\ref{FigWaveFunction} and remark that, in accordance with expectations, the momentum-space range of a ground-state baryon's Faddeev wave function increases as the current-mass of the valence constituents grows, an effect which corresponds to the system contracting in configuration space.  To be quantitative, we define $\ell_x^2$ to be the full width at half maximum of a given curve in Fig.\,\ref{FigWaveFunction} and associate a configuration space size via $s = 1/\ell_x$, using which conventions one finds:
\begin{subequations}
\begin{align}
s_{\Omega_{sss}} =0.85 \, s_{\Delta^{++}}\,,\\
s_{\Omega_{ccc}} =0.73 \, s_{\Delta^{++}}\,,\\
s_{\Omega_{bbb}} =0.41 \, s_{\Delta^{++}}\,.
\end{align}
\end{subequations}

Moving to Fig.\,\ref{FigWaveFunctionE}, it is evident that the same thing happens in the excited states, with approximately the same strength.  (\emph{N.B}.\, The $x$-axis scale in the lower panels covers a domain that is twice as large as that in the upper panels.)

\section{Epilogue}
\label{SecEpilogue}
We described a unified study of an array of mesons and baryons constituted from light- and heavy-quarks, using a symmetry-preserving rainbow-ladder truncation of the relevant bound-state equations in relativistic quantum field theory: the gap-, Bethe-Salpeter- and Faddeev-equations.  In particular, we computed the spectrum and leptonic decay constants of ground-state pseudoscalar- and vector-mesons: $q^\prime \bar q$ and $Q^\prime \bar Q$, with $q^\prime,q=u,d,s$, $Q^\prime,Q = c,b$; and the masses of $J^P=3/2^+$ $qqq$, $QQQ$ ground state baryons and their first positive-parity excitations.  Analysing the results, we showed that equal spacing rules provide sound estimates for masses and decay constants; and subsequently used this property to compute masses of $3/2^+$$ccb$, $cbb$ baryons, and a range of related single- and doubly-heavy baryons.  Within errors, the results agree with those obtained using lattice-regularised QCD.

We also analysed the internal structure of the ground and first positive-parity excited states of $qqq$, $QQQ$ baryons by studying their rest-frame three-body dressed-quark angular-momentum fractions and the pointwise behaviour of their Faddeev wave functions.  This revealed that each system has a complicated angular momentum structure.  For instance, the ground states are all primarily $S$-wave in nature, but each possesses $P$-, $D$- and $F$-wave components, with the $P$-wave fraction being large in the $u$ and $s$-quark states; and, somewhat surprisingly, the first positive-parity excitation in each channel has a large $D$-wave component, which grows with increasing current-quark mass, but also exhibits features consistent with a radial excitation.
Additionally, the pointwise behaviour of the Faddeev wave functions indicates that the configuration space extent of such bound states decreases as the mass of the valence-quark constituents increases.

On the strength of our results we judge that, in addition to its other known strengths, the coherent rainbow-ladder (RL) truncation of all relevant bound-state equations can produce realistic results for the masses and wave functions of $3/2^+$ baryons involving heavy quarks, both the ground-states and first positive-parity excitations.  This claim can be tested, \emph{e.g}.\ by computing decay rates and transition form factors involving these baryons, the matrix elements for which are straightforward to evaluate consistently within the RL truncation.  This work has begun.

An important next step is the kindred analysis of $1/2^+$ heavy-quark baryons, for which RL truncation should be realistic.
Naturally, $1/2^-$ and $3/2^-$ heavy-quark bound states are also of interest; but, based on experience with mesons, RL truncation is unlikely to be reliable in these odd-parity channels because it ignores some crucial effects driven by dynamical chiral symmetry breaking (DCSB).  Reliable predictions for such systems will require either extension to baryons of the DCSB-improved kernels used efficaciously in the meson sector -- a challenging task, or use of simpler, phenomenologically motivated changes to RL truncation, designed to mimic the impact of DCSB effects.  These efforts, too, are underway.

%


\acknowledgments
We are grateful for insights provided by D.~Binosi, C.~Chen, F.~Gao, V.~Mokeev, W.~Roberts and J. Segovia,
and for the hospitality of RWTH Aachen University, III.\,Physikalisches Institut B, Aachen, Germany.
Research supported by:
Forschungszentrum J\"ulich GmbH;
and U.S.\ Department of Energy, Office of Science, Office of Nuclear Physics, contract no.~DE-AC02-06CH11357.

\appendix
\setcounter{equation}{0}
\setcounter{figure}{0}
\setcounter{table}{0}
\renewcommand{\theequation}{\Alph{section}.\arabic{equation}}
\renewcommand{\thetable}{\Alph{section}.\arabic{table}}
\renewcommand{\thefigure}{\Alph{section}.\arabic{figure}}

\section{Faddeev amplitude}
\label{FAmplitude}
With $P=p_1+p_2+p_3$, we introduce two relative momenta:
\begin{eqnarray}
        q = \tfrac{1}{\surd 2}( p_1-p_2)
        \quad p = \tfrac{1}{\surd 6} (p_1+p_2-2p_3)\,,
\end{eqnarray}
which are momentum-space Jacobi coordinates for a three body system.  In terms of these variables, one can proceed to hyperspherical coordinates by defining a hyperradius and hyperangle:
\begin{equation}
	x = q^2 + p^2, \quad z_x = \frac{q^2-p^2}{q^2+p^2}\,.
\end{equation}
Notably, the hyperradius is a Poincar\'e-invariant average of the dressed-quark relative momenta:
\begin{equation}
	x = \tfrac{1}{3} [  (p_1-p_2)^2 + (p_2-p_3)^2 + (p_3-p_1)^2 ] \,.
\end{equation}

Given that the Faddeev amplitude carries four distinct spinor indices and must satisfy the positive-energy Rarita-Schwinger equation, it can be written as a direct product of two Dirac matrices
\begin{align}
{\Psi}^{\alpha_1 \alpha_2 \alpha_3,\delta }_{\mu}&(p,q,P)
= [D_{\nu}(p,q,P)]^{\alpha_1\alpha_2}
    [\underline{D}(p,q)\mathscr{P}_{\mu\nu}({P})]^{\alpha_3\delta} \nonumber \\
& =: D_{\nu}(p,q,P) \otimes \underline{D}(p,q) \mathscr{P}_{\mu\nu}(P)\,,
\end{align}
where the positive-energy Rarita-Schwinger projector is defined as ($\hat{P}^2 = 1$)
\begin{subequations}
\begin{align}
	\mathscr{P}_{\mu\nu}({P}) & =
    \Lambda_+(\hat{P}) \left( T_{\mu\nu}^P - \tfrac{1}{3} \gamma^T_\mu \gamma^T_\nu \right) \,,\\
	\Lambda_{\pm}(\hat{P})&=\tfrac{1}{2}(1\pm \gamma\cdot {\hat{P}})\,,
\end{align}
\end{subequations}
with $T_{\mu\nu}^P = \delta_{\mu\nu}-\hat{P}_\mu \hat{P}_\nu$ and $\gamma^T_\mu = T_{\mu\nu}^P \gamma_\nu$.

If one now defines three additional orthogonal, dimensionless momentum variables:
\begin{equation}
        r = \frac{\hat{p} - z_p\hat{P}}{\sqrt{1-z_p^2}},
\quad t=\frac{\hat{q} - z_q\hat{P}}{\sqrt{1-z_q^2}},
\quad s = \frac{t - z_0 r}{\sqrt{1-z_0^2}} \,,
\end{equation}
where $z_p = \hat{p}\cdot \hat{P}$, $z_q = \hat{q}\cdot \hat{P}$, $z_0=r\cdot t$, which have the properties $r^2=s^2=1$, $r\cdot \hat{P} = s\cdot \hat{P} = r\cdot s=0$, then the amplitude can be written
\begin{align}
{\Psi} & \!^{\alpha_1 \alpha_2 \alpha_3,\delta }_{\mu}(p,q,P)
 = \sum_{a,i,j;{\mathpzc s}=\pm} {\mathpzc f}_{aij{\mathpzc s}}(x,z_x,z_p,z_q,z_0) \nonumber \\
& \times
\left[ \Theta^a  \Gamma_\nu^i(r,s)  \Lambda_{\mathpzc s}(\hat{P}) \mathcal{C} \otimes \Theta^a \Gamma^j(r,s) \mathscr{P}_{\mu\nu}(\hat{P})
\right] \,, \label{GenPsi}
\end{align}
where the ${\mathpzc f}_{aij{\mathpzc s}}$ are scalar functions of five arguments, $\mathcal{C} = \gamma_4\gamma_2$ is the charge conjugation matrix, and
\begin{subequations}
\begin{align}
\Theta^a & = \{ \mathbf{1}, \gamma_5, \gamma^T_\beta, \gamma_5\gamma^T_\beta \}\,,\\
\Gamma^i(r,s) & =  \{\mathbf{1}, \gamma\cdot r, \gamma\cdot {s}, \gamma\cdot {r}\gamma\cdot {s} \}\,,\\
\Gamma^j_\nu(r,s) &= \{\gamma^T_\nu, r_\nu, s_\nu \} \otimes \Gamma(r,s) \,.
\end{align}
\end{subequations}

The sum in Eq.\,\eqref{GenPsi} involves $4 \times 4\times (3\times 4) \times 2 = 384$ terms, but only 128 are independent: the Faddeev amplitude has $4^4 = 256$ spinor degrees of freedom and only $256/2 = 128$ possess positive parity.  Including the positive-energy constraint, too, one arrives at $64$ independent terms in the sum.  Turning now to the Lorentz index, $\mu$, there are nominally four degrees of freedom.  However, the constraints on the Rarita-Schwinger spinor reduce this to just two.  Hence, one requires a basis with $64 \times 2 = 128$ elements to completely represent the Faddeev amplitude.  Having arrived at this understanding, and defining ${\mathpzc z}:= (x,z_x,z_p,z_q,z_0)$, $\hat{{\mathpzc z}}=(r,s,\hat{P})$, we write
\begin{equation}
\label{EqPsivarsigma}
{\Psi}^{\alpha_1 \alpha_2 \alpha_3,\delta }_{\mu}(p,q,P)
 = \sum_{n=1}^{128} {\mathpzc f}_{n}({\mathpzc z})
                                  [\mathsf{X}_n]^{\alpha_1\alpha_2\alpha_3,\delta}_{\mu}(\hat{{\mathpzc z}})\,,
\end{equation}
where the basis elements, $\{{\mathsf{X}_n}\}$, are those listed in Ref.\,\cite{SanchisAlepuz:2011jn}, Table~II, which satisfy
\begin{equation}
\tfrac{1}{8}{\rm tr} {\bar{\mathsf{X}}_n} {\mathsf{X}_{n^\prime}}
= \tfrac{1}{8}[\bar{\mathsf{X}}_n]^{\alpha_2 \alpha_1 \delta,\alpha_3}_{\mu} \;
   [{\mathsf{X}_{n'}}]^{\alpha_1\alpha_2\alpha_3,\delta}_{\mu} = \delta_{n n^\prime}\,,
\end{equation}
where
\begin{align}
\nonumber
[\bar{\mathsf{X}}_n]&\,  \!_\mu^{\alpha_2 \alpha_1 \delta,\alpha_3}(p,q,P) \\
& =
- \mathcal{C}_{\alpha_2 \alpha_2^\prime} \mathcal{C}_{\delta \delta^\prime}
[{\mathsf{X}}_n]_\mu^{\alpha_1^\prime \alpha_2^\prime \alpha_3^\prime, \delta^\prime}(-p,-q,-P)
\mathcal{C}^\dagger_{\alpha_1^\prime \alpha_1} \mathcal{C}^\dagger_{\alpha_3^\prime \alpha_3}.
\end{align}
Notably, when used in connection with the baryon's Faddeev wave function, Eq.\,\eqref{FaddeevWaveF}, these elements provide a straightforward connection to the rest-frame wave-function's partial-wave decomposition in terms of dressed-quark spin, $S$, and angular-momentum, $L$.

\section{Numerical procedure}
\label{AppNumerical}
A fully numerical approach to solving the Faddeev equation depicted in Fig.\,\ref{FEimage} presents a challenging exercise \cite{Eichmann:2011vu}.  In order to improve efficiency, we introduce incomplete wave functions:
\begin{align}
\nonumber
& _3 {\tilde\chi}^{\alpha_1\alpha_2\alpha_3,\delta}_{\mu}(p,q,P) =\sum_{n=1}^{128} \tilde{\mathpzc x}_n({\mathpzc z})\; [\mathsf{X}_n]_\mu^{\alpha_1\alpha_2\alpha_3,\delta}(\hat{{\mathpzc z}})\,;\\
& = S_{\alpha_1 \alpha_1^\prime}(p_1) S_{\alpha_2 \alpha_2^\prime}(p_2) {\Psi}^{\alpha_1\alpha_2\alpha_3,\delta}_{\mu}(p_1,p_2,p_3)\,,
\label{EqChiI}
\end{align}
with analogous definitions for $_{1,2}{\tilde\chi}$; and incomplete amplitudes:
\begin{subequations}
\begin{align}
_3 {\tilde\Psi} & \, \!^{\alpha_1\alpha_2\alpha_3,\delta}_{\mu}(p,q,P)
 =\sum_{n=1}^{128} {\mathpzc p}_n({\mathpzc z})\; [\mathsf{X}_n]_\mu^{\alpha_1\alpha_2\alpha_3,\delta}(\hat{{\mathpzc z}})\,;\\
& = \int_{dk} \mathscr{K}_{\alpha_1\alpha_1^\prime,\alpha_2\alpha_2^\prime}(k)\,
_3 \tilde{\chi}^{\alpha_1\alpha_2\alpha_3,\delta}_{\mu}(p,q_k,P)\,,
\end{align}
\end{subequations}
where $q_k = q - \surd 2\, k$, with $_{1,2}{\tilde\Psi}$ defined similarly, in terms of which the Faddeev equation, Eq.\,\eqref{eq:faddeev0}, can be rewritten
\begin{equation}
\label{eq:faddeev1}
\Psi^{\alpha_1\alpha_2\alpha_3,\delta}_{\mu}(p,q,P)
= \sum_{k=1,2,3}\! \,_k {\tilde\Psi}  \, \!^{\alpha_1\alpha_2\alpha_3,\delta}_{\mu}(p,q,P)
\end{equation}

We then compute projections using the orthonormal basis introduced in Appendix~\ref{FAmplitude} and thereby transform Eq.\,\eqref{eq:faddeev1} into a system of linear equations:
{\allowdisplaybreaks
\begin{subequations}
\label{eq:linfaddeev}
\begin{align}
\tilde{\mathpzc x}_n({\mathpzc z}) & = \sum_m \mathcal{G}_{nm} \psi_m({\mathpzc z})\\
%
 {\mathpzc p}_n({\mathpzc z}) & = \sum_m \int_{dk} \mathcal{K}_{nm} \tilde{\mathpzc x}_m({\mathpzc z}_k)\\
%
\nonumber
\psi_n({\mathpzc z}) & = {\mathpzc p}_n({\mathpzc z}) \\
  &  + \sum_m \mathcal{M}^{(1)}_{nm} {\mathpzc p}_m({\mathpzc z}_1)
    + \sum_m \mathcal{M}^{(2)}_{nm} {\mathpzc p}_m({\mathpzc z}_2)
\end{align}
\end{subequations}}
\hspace*{-0.5\parindent}where $\mathcal{G}$, $\mathcal{K}$, $\mathcal{M}^{(1,2)}$ are the following $128\times 128$ matrices:
{\allowdisplaybreaks
\begin{subequations}
\begin{align}
\nonumber
\mathcal{G}_{nm} & =
\tfrac{1}{8} [\bar{\mathsf{X}}_n]_\mu^{\alpha_2 \alpha_1 \delta,\alpha_3}(\hat{\mathpzc z})
    S_{\alpha_1\alpha_1^\prime}(p_1) S_{\alpha_2 \alpha_2^\prime}(p_2)\\
   & \quad \times  [\mathsf{X}_m]_\mu^{\alpha_1^\prime \alpha_2^\prime \alpha_3,\delta}(\hat{\mathpzc z}) \,, \\
\nonumber
\mathcal{K}_{nm} & =\tfrac{1}{8} [\bar{\mathsf{X}}_n]_\mu^{\alpha_2 \alpha_1 \delta,\alpha_3}(\hat{\mathpzc z})
 \mathscr{K}_{\alpha_1\alpha_1^\prime,\alpha_2\alpha_2^\prime}(k)\, \\
   & \quad \times  [\mathsf{X}_m]_\mu^{\alpha_1^\prime \alpha_2^\prime \alpha_3,\delta}(\hat{\mathpzc z}_k)  \,,\\
%
\mathcal{M}^{(1)}_{nm} & =
\tfrac{1}{8} [\bar{\mathsf{X}}_n]_\mu^{\alpha_2 \alpha_1 \delta,\alpha_3}(\hat{\mathpzc z})
                   [\mathsf{X}_m]_\mu^{\alpha_2 \alpha_3 \alpha_1,\delta}(\hat{\mathpzc z}_1) \,, \\
%
\mathcal{M}^{(2)}_{nm} & =
\tfrac{1}{8} [\bar{\mathsf{X}}_n]_\mu^{\alpha_2 \alpha_1 \delta,\alpha_3}(\hat{\mathpzc z})    [\mathsf{X}_m]_\mu^{\alpha_3 \alpha_1 \alpha_2,\delta}(\hat{\mathpzc z}_2)  \,.
\end{align}
\end{subequations}}

In Eqs.\,\eqref{eq:linfaddeev}, the unknown function is the full Faddeev amplitude $\psi_m({{\mathpzc z}})$, which can be solved iteratively. The intermediate matrices $\mathcal{G},\mathcal{K},\mathcal{M}^{(1)},\mathcal{M}^{(2)}$ may be stored in the memory to accelerate the iteration process.


\end{document}